\documentclass{elsarticle}

\usepackage{lineno}
\let\linenumbers\nolinenumbers\nolinenumbers
\journal{Journal of \LaTeX\ Templates}

\usepackage{graphicx}
\usepackage{multirow}

\usepackage{subcaption}
\usepackage{amsmath}
\usepackage{amssymb}
\usepackage{multibib}
\renewcommand{\cite}{\citep}
\setcitestyle{square}
\usepackage{appendix}
\usepackage{enumitem}
\usepackage{float}
\usepackage{changepage}
\usepackage{graphicx} 
\usepackage[export]{adjustbox}

\biboptions{authoryear}

\begin{document}

\begin{frontmatter}

\title{Exploring Gaussian mixture model framework for speaker adaptation of deep neural network acoustic models}

\author[mymainaddress]{Natalia Tomashenko\corref{mycorrespondingauthor}}

\author[mysecondaryaddress3]{Yuri Khokhlov}

\author[mymainaddress]{Yannick~Est\`eve}

\address[mymainaddress]{LIA, University of Avignon, France}

\address[mysecondaryaddress3]{STC-innovations Ltd, Saint-Petersburg, Russia}

\begin{abstract}
In this paper we investigate the GMM-derived (GMMD) features for adaptation of deep neural network (DNN) acoustic models. 
The adaptation of the DNN trained on GMMD features is done through the maximum a posteriori (MAP) adaptation of the auxiliary
GMM model used for GMMD feature extraction.
We explore fusion of the adapted GMMD features with conventional features, such as bottleneck and MFCC features, in two different neural network architectures: DNN and time-delay neural network (TDNN).
We analyze and compare different types of adaptation techniques such as i-vectors and feature-space adaptation techniques based on maximum likelihood linear regression (fMLLR) with the proposed adaptation approach, and explore their complementarity using various types of fusion such as feature level, posterior level, lattice level and others in order to discover the best possible way of combination.
Experimental results on the TED-LIUM corpus show that the proposed adaptation technique can be effectively integrated into DNN and TDNN setups at different levels and provide additional gain in recognition performance: up to $6\%$ of relative word error rate reduction (WERR) over the strong feature-space adaptation techniques based on maximum likelihood linear regression (fMLLR) speaker adapted DNN baseline, and up to $18\%$ of relative WERR in comparison with a speaker independent (SI) DNN baseline model, trained on conventional features. 
For TDNN models the proposed approach achieves up to $26\%$ of relative WERR in comparison with a SI baseline,  and up $13\%$  in comparison with the model adapted by using i-vectors. 
The analysis of the adapted GMMD features from various points of view demonstrates their effectiveness at different levels.
\end{abstract}

\begin{keyword}
Acoustic model adaptation \sep
Deep Neural Networks (DNN) \sep
Automatic Speech Recognition (ASR) \sep 
Gaussian Mixture Models (GMM) \sep 
Speaker adaptation \sep
GMM-derived (GMMD) features
\end{keyword}

\end{frontmatter}

\linenumbers

\section{Introduction}
Adaptation of DNN acoustic models is a rapidly developing  research area.
The aim of acoustic model (AM) adaptation is to reduce mismatches
between training and testing acoustic conditions and improve the accuracy of the automatic speech  recognition (ASR) system for a target speaker or channel using a limited  amount of adaptation data from the target acoustic source.
In the recent years DNNs have replaced conventional Gaussian mixture models (GMMs) in most state-of-the-art ASR systems, because it has been shown 
that DNN Hidden Markov Models (HMMs)  outperform GMM-HMMs in different ASR tasks. 
Many adaptation algorithms that have been developed for GMM-HMM systems \cite{gales1998maximum, gauvain1994maximum}
cannot be easily applied to DNNs because of the different nature of these models.
Among the adaptation methods developed for DNNs only a few take advantage of robust adaptability of  GMMs 
\cite{seide2011feature, rath2013improved, kanagawa2015feature, lei2013deep, liu2014combining, murali2015speaker, parthasarathi2015fmllr}.
However, none of them suggests a universal method for efficient transfer of all adaptation algorithms from the GMM models to DNN framework.

In the past, there were different attempts to integrate GMM and DNN models into a single structure. One of the common approaches is to use
features generated by neural networks, such as \textit{tandem}~\cite{hermansky2000tandem} or
\textit{bottle\-neck} (BN)~\cite{grezl2007probabilistic,yu2011improved,paulik2013lattice} features
in order to  train a GMM model.  
Other approaches include \textit{deep GMMs}~\cite{demuynck2013porting} and
a softmax layer with hidden variables~\cite{tuske2015integrating,tuske2015speaker}, which  use the concept of log-linear mixture models. 
In~\cite{variani2015gaussian}, a GMM layer is used as an alternative to the softmax layer in a DNN model.

In this paper we investigate a
recently introduced  GMM framework for adaptation of DNN-HMM acoustic models \cite{tomashenko2014speaker, tomashenko2015gmm, tomashenko2016ontheuse,Tomashenko2016anew,tomashenko2016exploration}. 
Our approach is based on using features derived from a GMM model for training DNN models \cite{tomashenko2014speaker, tomashenko2015gmm, tomashenko2016ontheuse, pinto2008combining} and GMM-based adaptation techniques.
In the previous works it was shown that  GMM log-likelihoods can be effectively used as features for training a DNN HMM model, as well as for the speaker adaptation task.

The first objective of this paper is to propose a universal way of integration of the GMM adaptation framework into the most commonly used neural network AMs, such as DNN (Section~\ref{DNN-GMMD}) and time delay neural network (TDNN) AMs (Section~\ref{TDNN-GMMD}) using MAP adaptation (Section~\ref{sec_map}) as an example.

Another important objective is to present an extensive experimental analysis of the proposed adaptation approach on the standard TED-LIUM corpus \cite{rousseau2014enhancing} for different types of neural network AMs.
These experiments include: 
adaptation of both cross-entropy (CE) and sequence trained DNN acoustic models (Section~\ref{sec_dnn_results});
adaptation of TDNN AMs (Section~\ref{sec_tdnn_results});
complementarity of the proposed approach with the two most popular adaptation techniques, such as  fMLLR (Section~\ref{sec_dnn_results}) and i-vectors (Section~\ref{sec_tdnn_results});
discovering the best possible way of information fusion (Section~\ref{Sect_Fus}) from the AMs trained with GMM-derived (GMMD) features and
the baseline conventional AMs, both for DNN and TDNN AMs, in order to improve the overall recognition accuracy.

The final goal is to look more deeply into the nature of the GMMD  features and adaptation techniques associated with them for better understanding their properties, strengths and weaknesses and the potential for improvement. For this purpose we perform a series of experiments on TDNN AMs (Section~\ref{sec_feature_analysis}) using lattice-based features (Section~\ref{sec_lattice-based-features}) 
by means of t-distributed stochastic neighbor embedding (t-SNE) visual analysis (Sections~\ref{sec_tsne}, \ref{sec_anal_tdnn}), 
using Davies-Bouldin (DB) index  (Sections~\ref{sec_db_inex}, \ref{sec_anal_tdnn}),
and different distributions for lattice-based features statistics.

\section{Review on neural network acoustic model adaptation}\label{Sec_Rev}

Various adaptation methods have been developed for
DNNs. 
These methods can  be categorized in  two broad classes, \textit{feature-space} and \textit{model-based} methods.

 \textit{Model-based adaptation} methods rely on direct modifications of DNN model parameters. 
 In~\cite{swietojanski2014learning, Swietojanski2016learning}, learning speaker-specific hidden unit contributions (LHUC) was proposed. 
The main idea of LHUC is to directly parametrize amplitudes of  hidden units, using a speaker-dependent amplitude function.
The idea of learning amplitudes of activation functions was also studied   in~\cite{trentin2001networks}.
The adaptation parameters estimation via maximum a posteriori (MAP) linear regression was proposed in~\cite{huang2014feature}, 
and a hierarchical MAP approach was studied in~\cite{huang2015maximum}. 
Other model-based DNN adaptation techniques include linear transformations, adaptation using regularization techniques,
subspace methods and others.

\textit{Feature-space adaptation} methods operate in the feature space and can either transform input features for DNNs, as it is done, for example, in fMLLR adaptation~\cite{seide2011feature} or use auxiliary features.

\subsection{Linear transformation}

One of the first adaptation methods developed for DNNs was 
linear transformation that can be applied at different levels of the DNN-HMM system: to the input features, as in linear input network transformation (LIN) \cite{neto1995speaker, gemello2006adaptation, li2010comparison}
 or feature-space discriminative linear regression (fDLR) 
 \cite{seide2011feature, yao2012adaptation};
 to the activations of hidden layers, as in linear hidden network transformation (LHN) \cite{gemello2006adaptation}; 
or to the softmax layer, as in 
LON \cite{li2010comparison} or in output-feature discriminative linear regression \cite{yao2012adaptation}. 

\subsection{Regularization techniques}

In order to  improve generalization during the adaptation, regularization techniques,  
such as L2-prior regularization~\cite{liao2013speaker}, 
Kullback-Leibler divergence regularization~\cite{yu2013kl, huang2015regularized, toth2016adaptation}
conservative training~\cite{albesano2006adaptation}
and others~\cite{ochiai2014speaker} are used. 
\subsection{Multi-task learning}

The concept of multi-task learning (MTL) has recently been applied to the task of speaker adaptation in several works  
\cite{li2015ensemble, huang2015rapid, swietojanski2015structured}
and has been shown to improve the performance of different model-based DNN adaptation techniques, 
such as LHN \cite{huang2015rapid} and LHUC \cite{swietojanski2015structured}. 
A slightly different idea was proposed earlier in \cite{Price2014hierarchy} in the form of
special hierarchy of output layers, where tied triphone states are followed by monophone states.

\subsection{Subspace methods}
Subspace adaptation methods  aim to 
find a speaker subspace and then construct the adapted
DNN parameters as a point in the subspace.
In~\cite{dupont2000fast} an approach similar to 
the eigenvoice technique~\cite{kuhn2000rapid}, was proposed for the fast speaker adaptation of neural network AMs.

In~\cite{wu2015multi} a \textit{multi-basis adaptive neural network} is proposed, where
a traditional DNN topology
is modified 
and a set of sub-networks, referred as \textit{bases} were introduced. 
This DNN has a common input layer and 
a common output layer for all the bases.
Each basis has several fully-connected hidden layers and there is no connections between neurons from different bases.
The outputs of bases are combined
by linear interpolation using a set of adaptive weights. The adaptation to a given speaker can be  performed through optimization of interpolation weights for this speaker. 
The idea of this approach was motivated by the cluster adaptive training (CAT), developed for GMM AMs.
Paper~\cite{tan2015cluster} also investigates  the CAT framework for DNNs.
A subspace learning speaker-specific
hidden unit contributions (LHUC) adaptation was proposed in~\cite{samarakoon2016subspace}.

\subsection{Factorized adaptation}

Factorized adaptation \cite{li2014factorized,yu2012factorized,qian2016neural,tran2016factorized,samarakoon2016multi} takes into account different factors that influence the speech signal.
These factors can have different nature (speaker, channel, background noise conditions and others) and can be modeled explicitly before incorporating them
into the DNN structure, for example, in the form of auxiliary features~\cite{li2014factorized}, such as i-vectors, or can be learnt jointly with the neural network AM. 
The first case, when factors, such as noise or speaker information, are estimated explicitly from the training and testing data, and are then fed to the DNN AM, is also known as 
\textit{noise-aware} or \textit{speaker-aware training} correspondingly~\cite{yu2014automatic}.
In paper~\cite{yu2012factorized} 
two types of factorized DNNs were introduced: \textit{joint and disjoint models}.
In~\cite{tran2016factorized} an extension of the LIN adaptation, so-called \textit{factorized LIN} (FLIN), has been investigated for the case when adaptation data for a given speaker include multiple acoustic conditions. The feature transformations are represented as weighted combinations of affine transformations of the enhanced input features.

\subsection{Auxiliary features}\label{sec_ivect}
Using auxiliary features, such as i-vectors 
\cite{saon2013speaker,karanasou2014adaptation, 
gupta2014vector, 
senior2014improving},
is another widely used approach in which the acoustic feature vectors are augmented with additional 
speaker-specific or channel-specific features computed for each speaker or utterance at both training and test stages. 
Originally i-vectors were developed for speaker verification and speaker recognition tasks~\cite{dehak2011front},
and nowadays they have become a very common technique in these domains. 
I-vectors can capture the relevant information about the speaker in a low-dimensional fixed-length representation~\cite{dehak2011front}. They were first applied for adaptation of GMM-HMM models~\cite{karafiat2011ivector}, and later for DNN-HMMs~\cite{saon2013speaker, senior2014improving, karanasou2014adaptation, gupta2014vector}.
Another example of auxiliary features is the use
of speaker-dependent bottleneck (BN) features obtained from a speaker
aware DNN used in a far field speech recognition task~\cite{liu2014using}.
Alternative methods include adaptation with speaker codes~\cite{Abdel-Hamid2013fast, xue2014fast}.

We will describe i-vectors in more details here because this is one of the most popular methods for DNN adaptation, and we will use this technique as a baseline for comparison with the proposed approach in our experiments in
Section~\ref{sec_tdnn_results}.

\subsubsection*{I-vector extraction}
The acoustic feature vector $\textbf{o}_t\in\mathbb{R}^d$ can be considered as a sample, generated with a \textit{universal background model} (UBM), represented as a GMM with K diagonal covariance Gaussians~\cite{dehak2011front,saon2013speaker}:
\begin{equation}\label{eq:ubm}
\textbf{o}_t\sim \sum_{k=1}^{K}c_k\mathcal{N}\left(\cdot;\pmb{\mu}_k,\pmb{\Sigma}_k\right),
\end{equation}
where $c_k$ are the mixture weights, $\pmb{\mu}_k$ are means and $\pmb{\Sigma}_k$ are diagonal covariances. The acoustic feature vector $\textbf{o}_t(s)$, belonging to a  given speaker $s$ is described with the distribution:
\begin{equation}\label{eq:ivect}
\textbf{o}_t(s)\sim \sum_{k=1}^{K}c_k\mathcal{N}\left(\cdot;\pmb{\mu}_k(s),\pmb{\Sigma}_k\right),
\end{equation}
where $\pmb{\mu}_k(s)$ are the  means of the GMM, adapted to the speaker $s$.
It is assumed that there is a linear dependence between the speaker-dependent (SD) means $\pmb{\mu}_k(s)$  and the speaker-independent (SI) means $\pmb{\mu}_k$, which can be expressed in the form:
\begin{equation}
\pmb{\mu}_k(s)=\pmb{\mu}_k+\textbf{T}_k\textbf{w}(s),~~~k=1,\ldots,K,
\end{equation}
where $\textbf{T}_k\in \mathbb{R}^{D\times M}$ is a \textit{factor loading matrix}, corresponding to component $k$ and 
i-vector corresponding to speaker $s$ is estimated as the mean of the
 distribution of  $\textbf{w}(s)$.
Each $\textbf{T}_k$ contains $M$ bases,
that span the subspace of the important variability in the component mean vector space, corresponding to component~$k$.

The detailed description of how to estimate the factor loading matrix, given the training data $\{\textbf{o}_t\}$, and how to estimate i-vectors $\textbf{w}(s)$, given $\textbf{T}_k$ and speaker data $\{\textbf{o}_t(s)\}$, can be found, for example, in~\cite{dehak2011front,saon2013speaker}.

\subsubsection*{Integration of i-vectors into a DNN model}
Various methods of i-vector integration into a DNN AM have been proposed in the literature.

The most common approach \cite{saon2013speaker, senior2014improving, gupta2014vector} is to estimate  i-vectors for each speaker (or utterance), and then to concatenate it with acoustic feature vectors, belonging to a corresponding speaker (or utterance). The obtained concatenated vectors are introduced to a DNN for training.
In the test stage i-vectors for test speakers also have to be estimated, and input in a DNN in the same manner.

Unlike acoustic feature vectors, which are specific for each frame, 
an i-vector is the same for a chosen group of
acoustic features, to which it is appended.
For example, i-vector can be calculated for each utterance, as in~\cite{senior2014improving}, or estimated using all the data of a given speaker, as in~\cite{saon2013speaker}. 
I-vectors encode those effects in the acoustic signal, to which an ASR system is desired to be invariant: speaker, channel and background noise.
Providing to the input of a DNN the information about these factors makes it possible for a DNN to 
normalize the acoustic signal with respect to them.

An alternative approach of i-vector integration into the DNN topology is presented in 
\cite{miao2015speaker, miao2014towards}, where an input acoustic feature vector is normalized  though a linear combination of it with a speaker-specific normalization vector obtained from an i-vector.
Similar approaches have been studied in~\cite{lee2016semisupervised,goo2016speaker}.
Also i-vector dependent feature space transformations were proposed in~\cite{li2015vector}.

\subsection{Adaptation based on GMMs}

The  idea of integrating  generative models into  discriminate classifiers is not new. In the past, one solution to this problem was to use so-called \textit{kernels} that are calculated using generative models~\cite{jaakkola1999exploiting,longworth2009combining,gales2010discriminative,ragni2011derivative}.
Kernel methods were designed to allow  classifiers (such as  support vector machines (SVMs) as in~\cite{gales2010discriminative,longworth2009combining,smith2002speech}) to handle sequential data and to map variable length data sequences into a fixed dimensional representation. 
There are several papers devoted to these approaches in different domains, for example, for biosequence analysis~\cite{jaakkola1999exploiting},
speaker verification~\cite{longworth2009combining},
and for noise robust speech recognition~\cite{gales2010discriminative,ragni2011derivative}. 
Using generative models to compute kernels also allows  to use compensation and adaptation techniques for classifiers through the adaptation of generative kernels~\cite{gales2010discriminative}.

The most common way of combining GMM and DNN models for adaptation is using GMM-adapted features,
for example fMLLR, as input for DNN training~\cite{seide2011feature, rath2013improved, kanagawa2015feature, parthasarathi2015fmllr}. In~\cite{lei2013deep} likelihood
scores from DNN and GMM models, both adapted in the
feature space using the same fMLLR transform, are combined
at the state level during decoding.
Similar ideas are also presented in~\cite{swietojanski2013revisiting}.
Other methods include  temporally varying weight regression~\cite{liu2014combining} and
GMMD features~\cite{tomashenko2014speaker, tomashenko2016exploring, tomashenko2015gmm}.

\section{Hybrid DNN-HMM systems with GMMD features}\label{Sect_SAT}

In a conventional GMM-HMM ASR system, the state emission log-likelihood of the observation feature vector $\textbf{o}_t$ at time $t$ 
for certain tied state $s_i$ of HMMs is modeled as
\begin{equation}\label{eq:log-like}
\log{p(\textbf{o}_t\mid s_i)}=\log{\sum_{m=1}^{M_i}{w_{im}\mathcal{N}(\textbf{o}_t; \pmb{\mu}_{im},\pmb{\Sigma}_{im})}},
\end{equation}
where $M_i$ is the number of Gaussian distributions in the GMM for state $s_i$; $w_{im}$ is the  mixture weight of the m'th component of in the mixture for state $s_i$; 
$\pmb{\mu}_{im}$ is the corresponding mean vector, and $\pmb{\Sigma}_{im}$ is  the  covariance matrix.

In a DNN-HMM system, outputs of a DNN are the state posteriors $p(s_i|\textbf{o}_t)$,
which are transformed for decoding into pseudo (or scaled) likelihoods as follows
\begin{equation}
p(\textbf{o}_t\mid s_i)=
\frac{p(s_i\mid \textbf{o}_t)p(\textbf{o}_t)}{p(s_i)}\propto
\frac{p(s_i\mid \textbf{o}_t)}{p(s_i)},
\end{equation}
where the state prior ${p(s_i)}$ can be estimated from the state-level forced alignment on the training speech data,
and probability $p(\textbf{o}_t)$ is independent on the HMM state and can be omitted during the decoding process. Hence, log-likelihoods $\log{p(\textbf{o}_t\mid s_i)}$ can be estimated as $\log{p(s_i\mid \textbf{o}_t)}-\log{p(s_i)}$.

The use of log-likelihoods from a GMM model for training an MLP recognizer was investigated in~\cite{pinto2008combining}.
Construction of GMMD features for adapting DNNs was proposed in~\cite{tomashenko2014speaker, tomashenko2015gmm, tomashenko2016ontheuse}, where it was demonstrated, using MAP and fMLLR adaptation as an example, that this type of features provide 
a solution for efficient transferring GMM-HMM adaptation algorithms into the DNN framework. 
The same idea of using adapted GMMD features as input to DNNs was later applied  
to the task of noise adaptation~\cite{Kundu+2016}.

The GMMD features can be used directly to train DNN acoustic models, as  in~\cite{tomashenko2014speaker, tomashenko2015gmm}, or  in combination with other conventional features. 
In this paper, we present incorporation of the adapted GMMD features into several state-of-the-art recipes for neural network AM training. 

\subsection{Training DNN acoustic model with GMMD features}\label{DNN-GMMD}

The scheme for training DNN models with GMM adaptation framework is shown in Figure 1. 

\begin{figure}
\centering
    \includegraphics[width=74mm]{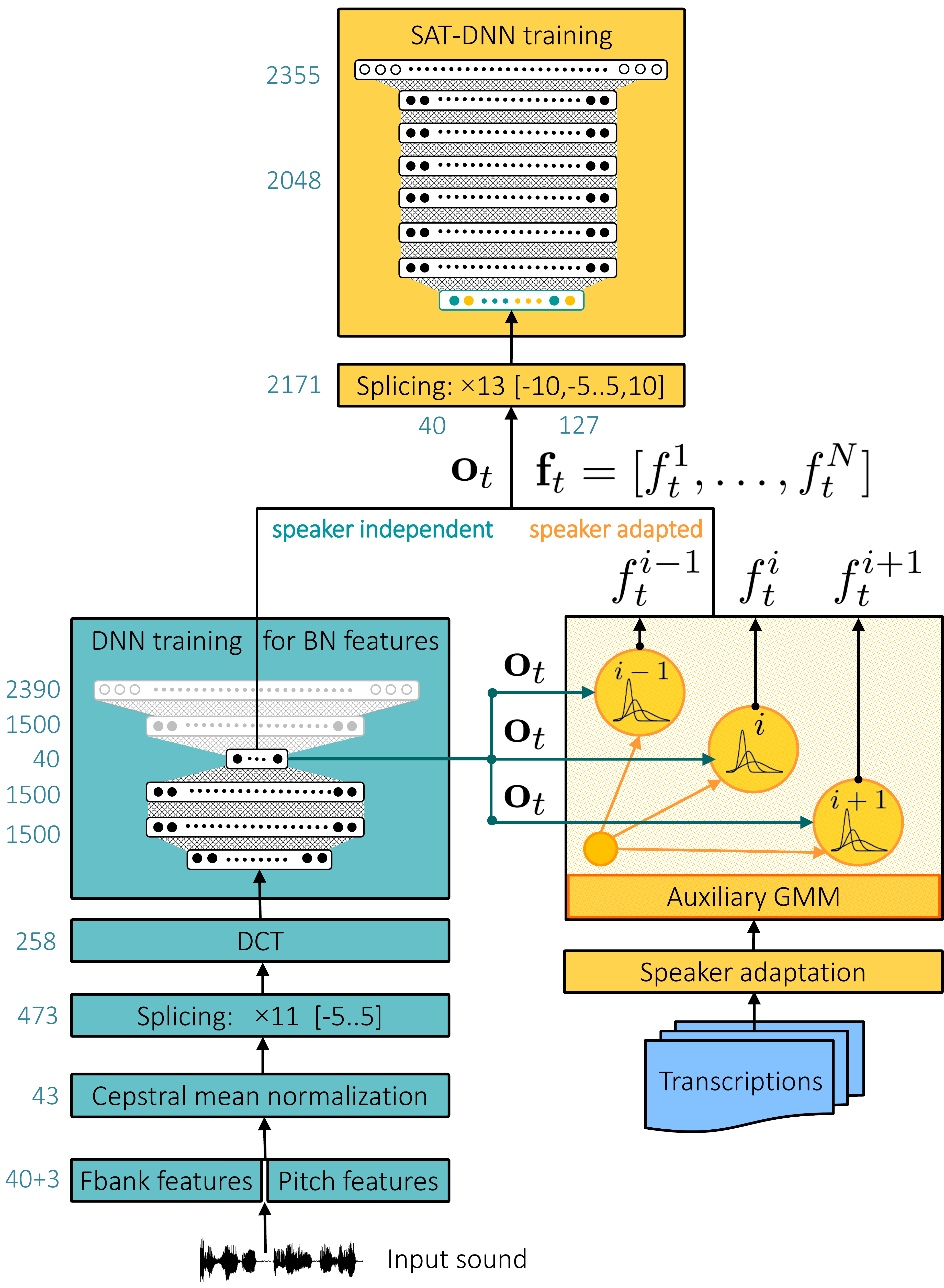}
	\caption{ {Using speaker adapted BN-based GMMD features for speaker adaptive training (SAT) of a  DNN-HMM.}}
	\label{SAT}
\end{figure}

First, 40-dimensional log-scale filterbank features, concatenated with 3-dim\-ensional pitch-features, are spliced across 11 neighboring frames (5 frames on each side of the current frame), resulting in 473-dimensional $(43\times11)$  feature vectors. After that, a DCT transform is applied and the dimension is reduced to 258. Then a DNN model for 40-dimensional bottleneck (BN) features is trained on these features.
An auxiliary triphone or monophone GMM model is used to transform BN feature vectors into log-likelihoods vectors. At this step, speaker adaptation of the auxiliary speaker-independent (SI) GMM-HMM  model is performed for each speaker in the training corpus and a new speaker-adapted (SA) GMM-HMM model is created in order to obtain SA GMMD features.

 For a given BN feature vector $\textbf{o}_t\in\mathbb{R}^d$, a new GMMD feature  vector $\textbf{f}_t$ is obtained by calculating log-likelihoods across all the states of the auxiliary GMM  model on the given vector as follows:
\begin{equation}
\textbf{f}_t = [f_t^1,\ldots,f_t^N],
\label{eq1}
\end{equation}
where $N$ is the number of states in the auxiliary GMM model,

\begin{equation}
f_t^i=\log{\left(p(\textbf{o}_t \mid s_t=i)\right)}
\label{eq2}
\end{equation}
is the log-likelihood estimated using the GMM. Here $s_t$ denotes the state index at  time $t$.
Formula~(\ref{eq2}) for the $i$-th component of the GMMD feature vector $\textbf{f}_t$ can be rewritten (using the notations from Formula~(\ref{eq:log-like})) as follows:
\begin{equation}
f_t^i=\log{\sum_{m=1}^{M_i}\frac{w_{im}}{\sqrt{(2\pi)^d|\pmb{\Sigma}_{im}|}}\exp\left\{-\frac{1}{2}(\textbf{o}_t-\pmb{\mu}_{im})^T\boldsymbol{\pmb{\Sigma}}_{im}^{-1}(\textbf{o}_t-\pmb{\mu}_{im})\right\}}.
\end{equation}

The obtained GMMD feature vector $\textbf{f}_t$ is concatenated with the original vector $\textbf{o}_t$. 
After that, the features are spliced in time taking a context size of 13 frames: [-10,-5..5,10]\footnote{The notation [-10,-5..5,10] means that for the current acoustic vector $\textbf{o}_t$, in the DNN training we use a context vector which consists of the following 13 frames:  $\left\{\textbf{o}_{t-10}, \textbf{o}_{t-5}, \textbf{o}_{t-4},\ldots,\textbf{o}_{t},\ldots\textbf{o}_{t+4},\textbf{o}_{t+5},\textbf{o}_{t+10}\right\}$.}.
These features are used as the input for speaker adaptive training (SAT) of a DNN. 
The proposed approach can be considered a feature space transformation technique with respect to DNN-HMMs trained on GMMD features.

Note, that the proposed GMMD features are very different from i-vectors (Section~\ref{sec_ivect}) in several aspects despite the fact that  both these methods are GMM-related.

First, i-vectors represent the acoustic characteristics of the speaker with respect to the general speaker distribution, which is characterized by a UBM (Formulas~(\ref{eq:ubm}),(\ref{eq:ivect})). GMMD features, when they are adapted, represent the speaker-adapted distributions of  acoustic classes. The better a GMMD vector is adapted, the closer these distributions are to speaker-dependent ones, and the higher  likelihoods for the acoustic vectors of a corresponding  speaker.

Second, i-vectors do not distinguish between acoustic classes, while in computation of GMMD features these information is explicitly represented by different components of  GMMD feature vectors. 
Each component of a GMMD feature vector is adapted to more closely match to the pronunciation of a given speaker of the corresponding acoustic class.

Also, since we have more classes in GMMD features to adapt individually, this means, that in comparison with i-vectors, they can potentially benefit more from adaptation than the  amount of adaptation data increases, especially when we use such adaptation techniques for GMMs as MAP.

Finally, i-vectors are usually computed for a sequences of vectors (per speaker, per utterance, or for a shorter time interval), while a GMMD feature vector is unique for each speech frame.

All these differences can also  allow us to suggest that both approaches can be complementary to each other. We will experimentally explore this question  in Section~\ref{Sect_Exp}.

\subsection{Training TDNN acoustic model with GMMD features}\label{TDNN-GMMD}

\begin{figure}[ht]
\centering
    \includegraphics[width=72mm]{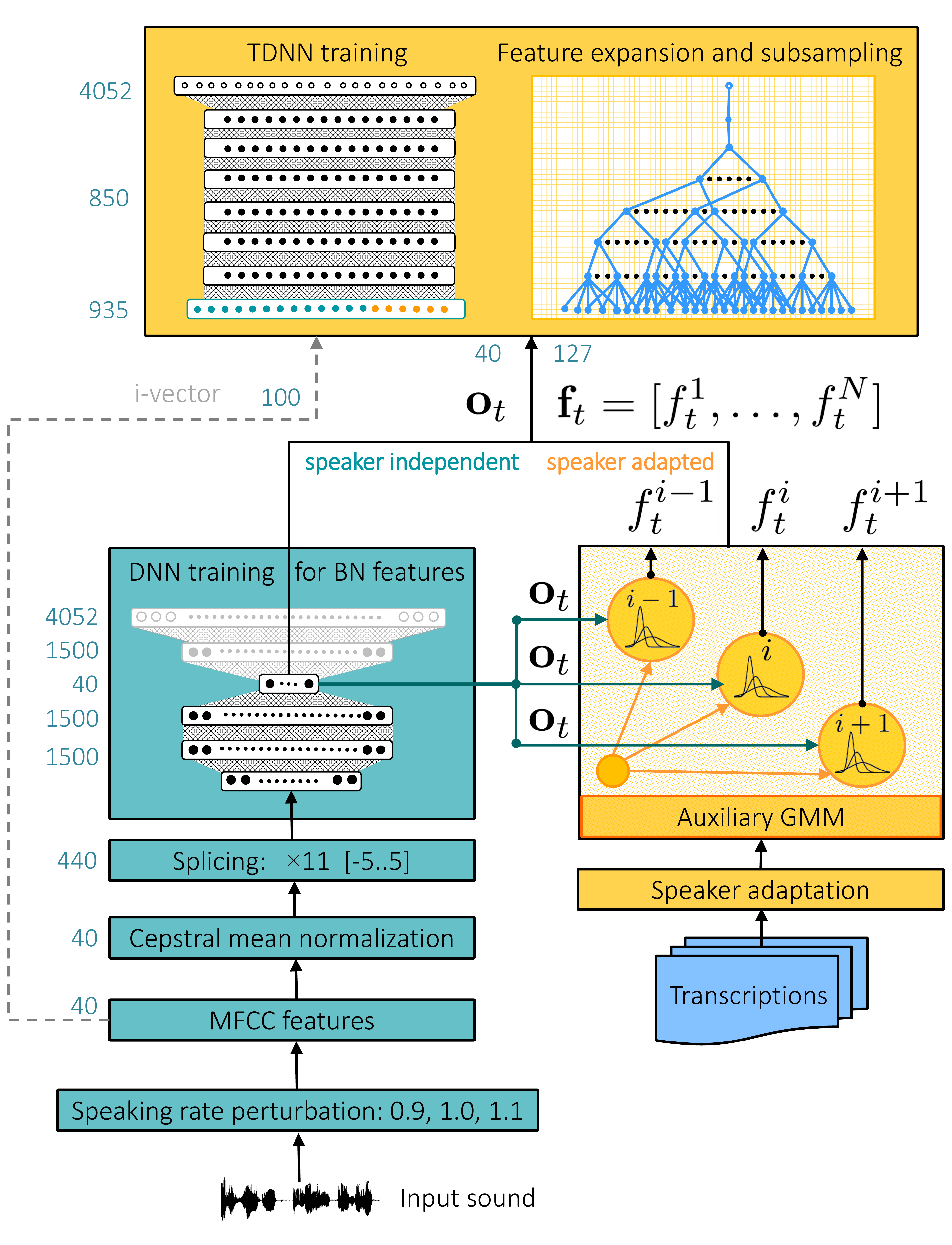}
	\caption{ {Using speaker adapted BN-based GMMD features for SAT TDNN training.}}
	\label{SAT_TDNN}
\end{figure}

In addition to the system described in Section \ref{DNN-GMMD}, we aim to explore the effectiveness of using GMMD features to train a time delay neural network (TDNN)~\cite{waibel1989phoneme}. 
A TDNN model architecture allows to capture the long term dependencies in speech signal. The recently proposed  approaches to train TDNN acoustic models~\cite{peddinti2015time} are reported to show higher performance on different LVCSR tasks compared with the standard (best) DNN systems.
We aim to incorporate GMMD features into the existing state-of-the art recipe for TDNN models~\cite{peddinti2015time}. For comparison purposes, we take a Kaldi TED-LIUM recipe with a TDNN acoustic model as a basis.
An example of using GMMD features for training a TDNN is shown in Figure~\ref{SAT_TDNN}. Here, as before, we use BN features to train the GMM auxiliary model for GMMD feature extraction. Then, GMMD features are obtained in the same way as described in Section~\ref{DNN-GMMD}.

There are several options for obtaining the final features, which are fed to the TDNN model. 
 GMMD features can be combined with the original MFCCs  or with BNs, that are used for training the auxiliary GMM model, as shown in Figure~\ref{SAT_TDNN}. In both cases, we can also use speaker i-vectors as complementary auxiliary features. 
All these possibilities will be  explored in  Section~\ref{sec:experim_tdnn}.

\subsection{MAP adaptation}
\label{sec_map}
In this work, we use the MAP adaptation algorithm~\cite{gauvain1994maximum} in order to adapt the SI GMM model.
Speaker adaptation of a DNN-HMM model built on GMMD features is performed through the MAP adaptation of the auxiliary GMM  model which is used for calculating GMMD features. 
Let $m$ denote an index of a Gaussian in the SI AM, and $\pmb{\mu}_m$ the mean of this Gaussian. Then the MAP estimation of the mean vector is 
\begin{equation}
\widehat{\pmb{\mu}}_m=\frac{\tau\pmb{\mu}_m+\sum_t{\gamma_m(t)  \textbf{o}_t}}{\tau+\sum_t{\gamma_m(t)}},
\label{eqmap}
\end{equation}
where $\tau$ is the parameter that controls the balance between the maximum likelihood estimate of the mean and its prior value;
$\gamma_m(t)$ is the posterior probability of Gaussian component $m$ at time $t$.

\section{Experimental study}\label{Sect_Exp}

\subsection{Data sets}
The experiments were conducted on the TED-LIUM corpus~\cite{rousseau2014enhancing}.
We used the last (second) release of this corpus. This publicly available data set contains 1495 TED talks that amount to 207 hours (141 hours of male, 66 hours of female) speech data from 1242 speakers, 16kHz.
For experiments with SAT and adaptation we removed from the original corpus data for those speakers, who had less than 
 5 minutes of data, and  from  the rest of the corpus we made four data sets:  
training set,  development set and two test sets. 
Characteristics of the obtained data sets are given in Table \ref{sets}.

\begin{table}
\caption{\label{sets}{Data sets statistics}}
\centering
		\begin{tabular}{|c|c|c|c|c|c|}
			\hline
			 \multicolumn{2}{|c|}{\multirow{2}{*}{Characteristic}}  & \multicolumn{4}{|c|}{Data set} \\ \cline{3-6}
			 \multicolumn{2}{|c|}{} &  Training & Development & $\mathrm{Test_1}$ & $\mathrm{Test_2}$ \\
			\hline   
			\multirow{3}{*}{\shortstack[c]{Duration, \\ hours}} &  Total & 171.66    & 3.49 & 3.49 & 4.90 \\
			&Male           & 120.50   & 1.76  & 1.76  & 3.51 \\
			&Female          & 51.15    &  1.73  & 1.73  & 1.39 \\ \hline 
			\multirow{3}{*}{\shortstack[c]{Duration \\ per speaker, \\ minutes}} & Mean   &  10.0 &  15.0 &  15.0 &  21.0    \\ 
			& Minimum &  5.0  &  14.4  & 14.4   &  18.3 \\ 
			& Maximum  &  18.3 &  15.4  & 15.4   & 24.9 \\ \hline 
			\multirow{3}{*}{\shortstack[c]{Number \\ of speakers}} & Total  & 1029    & 14  & 14  &  14 \\
			& Male                & 710     & 7  & 7  & 10 \\
			& Female              & 319     &  7  & 7  & 4 \\ \hline 
			Number of words & Total & - & 36672 & 35555 & 51452 \\ \hline
			\end{tabular}
\end{table}

For evaluation, two different language models (LMs) are used:
\begin{itemize}
	\item \textit{LM-cantab} is  a publicly available 3-gram language model 
\textit{cantab-TEDLIUM-pruned.lm3}\footnote{http://cantabresearch.com/cantab-TEDLIUM.tar.bz2} with 150K word vocabulary. The same LM was used in experiments presented in \cite{Tomashenko2016anew} and  in the Kaldi \textit{tedlium s5} recipe. 
  \item \textit{LM-lium}
is a 4-gram LM from TED-LIUM corpus with 152K word vocabulary, which is currently used in the Kaldi \textit{tedlium s5\_r2}   recipe.  
We conducted part of the experiments presented here using this LM in order to be compatible with the most recent Kaldi recipe and for comparison purposes with the results of the  TDNN acoustic models	
\end{itemize}

\subsection{System fusion}\label{Sect_Fus}

\begin{figure}[t]
\begin{center}
\begin{subfigure}{1\textwidth}
\center{\includegraphics[width=116mm]{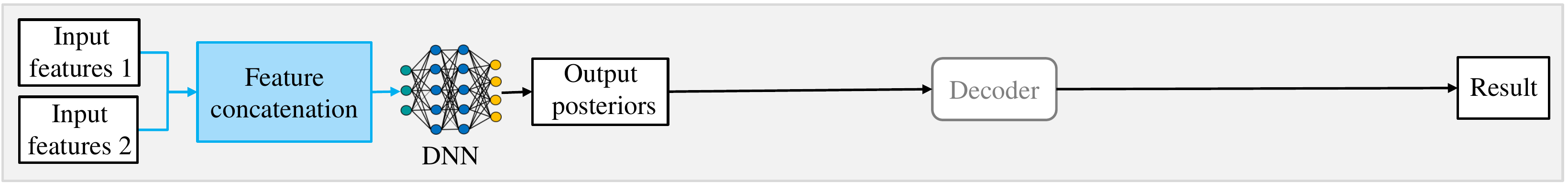}}
\caption{Fusion for training and decoding stages.}
\label{Fusion1}
\end{subfigure}
\begin{subfigure}{1\textwidth}
\center{\includegraphics[width=116mm]{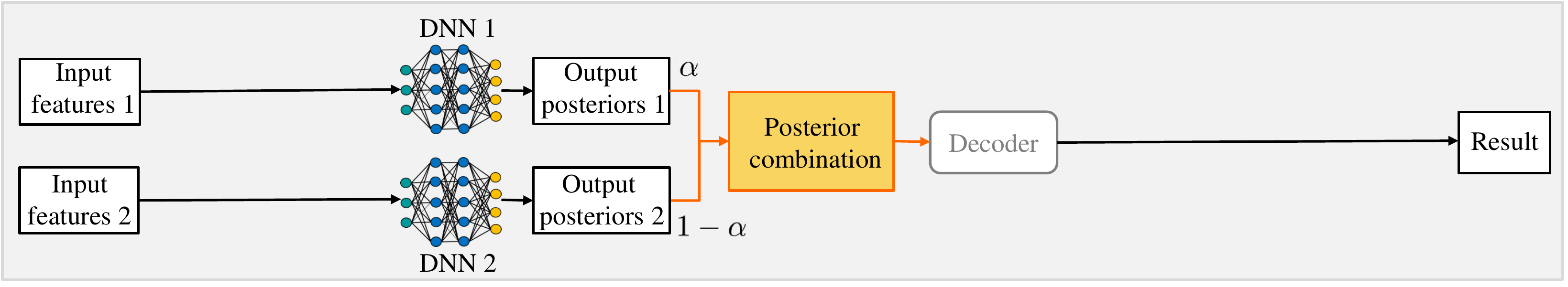}}
\caption{Fusion for decoding stage: posterior combination.}
\label{Fusion2}
\end{subfigure}
\begin{subfigure}{1\textwidth}
\center{\includegraphics[width=116mm]{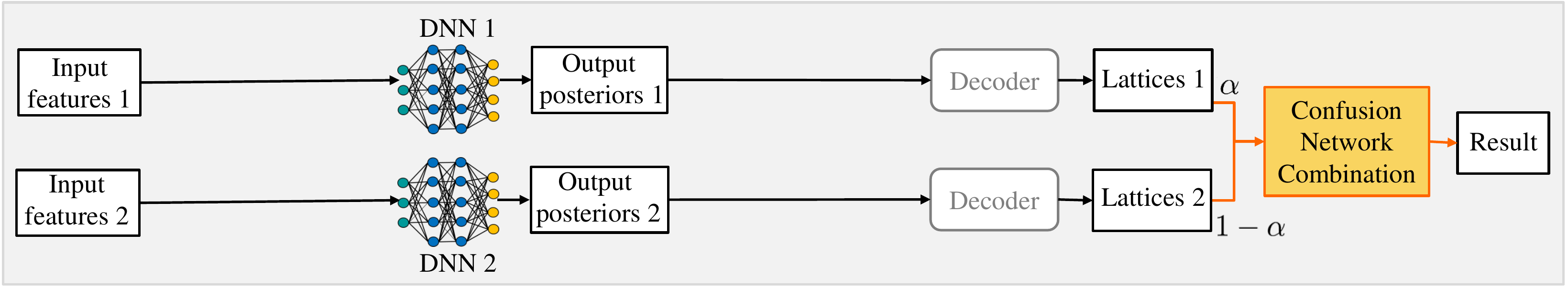}}
\caption{Fusion for decoding stage: CNC.}
\label{Fusion3}
\end{subfigure}
\begin{subfigure}{1\textwidth}
\center{\includegraphics[width=116mm]{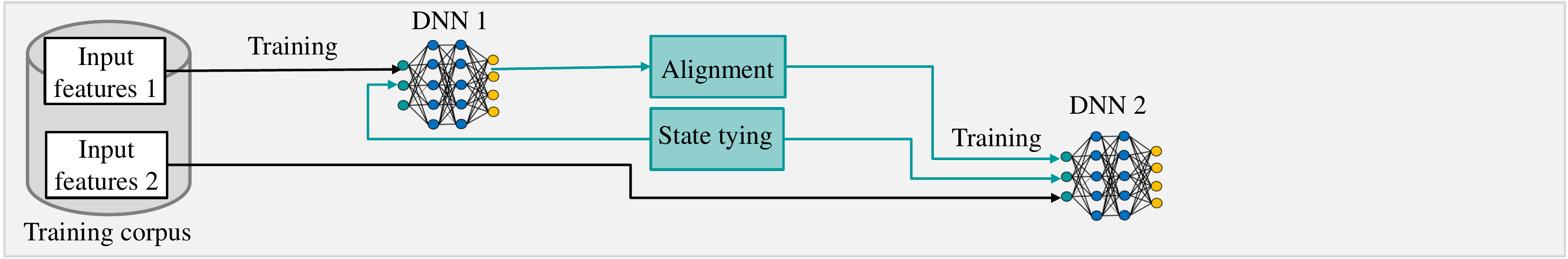}}
\caption{Fusion for training stage.}
\label{Fusion4}
\end{subfigure}
\end{center}
\caption{Types of fusion}
\end{figure}

In this section, we introduce several types of combination of GMMD features with conventional ones at different levels of DNN architecture.
It is known that GMM and DNN models can be complementary and their combination allows to improve the performance of ASR systems~\cite{pinto2008combining, swietojanski2013revisiting}. 
We explore the following types of fusion:
\begin{itemize}[wide=0pt]
\item \textit{Feature level fusion} (Figure~\ref{Fusion1}), where
input features are combined before performing classification~\cite{pinto2008combining}.
 In our case, features of different types -- GMMD  and cepstral or BN features are simply concatenated and provided as input into the DNN model for training.
This type of fusion allows us to combine different adaptation techniques in a single DNN model.
\item \textit{{Posterior (state) level fusion}} (Figure \ref{Fusion2}), where the outputs of two or more 
DNN models are combined on a state level~\cite{parthasarathi2015fmllr,pinto2008combining,lei2013deep,swietojanski2013revisiting}.
In this work, we perform frame-synchronous fusion using a linear combination of the observation log-likelihoods 
of two models ($\mathrm{DNN_1}$ and $\mathrm{DNN_2}$) as follows:
\begin{equation}
\log{(p(\textbf{o}_t \mid s_i))}=\alpha\log(p_{_{\mathrm{DNN_1}}}(\textbf{o}_t \mid s_i))+(1-\alpha)\log(p_{_\mathrm{DNN_2}}(\textbf{o}_t \mid s_i)),
\label{eq_fus}
\end{equation}
where $\alpha\in[0,1]$ is a weight factor that is optimized on a development set.
This approach assumes that both models have the same state tying structure. 
\item \textit{Lattice level fusion} (Figure~\ref{Fusion3}) is the highest level of fusion operates in the space of generated word hypotheses~\cite{fiscus1997post,evermann2000posterior}. In this work, we experiment with the Confusion Network Combination (CNC)~\cite{evermann2000posterior} approach, where confusion networks built from individual lattices are aligned.

\end{itemize}

There are other possible ways of  combining information from different ASR systems. 
In this paper, we used phoneme-to-speech alignment obtained by one acoustic DNN model to train another DNN model (Figure \ref{Fusion4}). 
In addition, we used state tying from the first DNN model to train the second DNN. 
This procedure is important when we want to apply \emph{posterior fusion} for two DNNs and need the same state tying for these models.

\subsection{Overview of experiments and questions addressed in the study}

We used the open-source Kaldi toolkit~\cite{povey2011kaldi} and mostly followed  the standard TED-LIUM Kaldi recipes to train the two baseline systems, 
 corresponding to two different types of acoustic models (DNN and TDNN)\footnote{using \textit{"nnet1"} and \textit{"nnet3"} Kaldi setups: http://kaldi-asr.org/doc/dnn.html}.

The following questions have been addressed in this section.

First, in order to  explore and prove the universality of the proposed adaptation approach,  we start our study from the classical fully connected DNN topology, and then choose TDNNs as one of the most efficient neural network architecture for AMs.

Second, we are interested in comparing the proposed adaptation technique with the two most popular adaptation approaches for neural network AMs: fMLLR (experiments for DNNs) and i-vectors (experiments for TDNNs).

Third, we study different types of fusion (features-, posterior-, lattice-level) of GMMD features with other features (both SI and adapted ones) for DNNs and TDNNs.
We  aim to explore the complementary of the proposed adaptation technique to other adaptation approaches: fMLLR (for DNNs) and i-vectors (for TDNNs).

Finally, the impact of the training criterion on the adaptation performance is investigated.

\subsection{Experiments with DNN models}\label{sec:experim_dnn}

This section presents systems and adaptation results for DNN AMs.

\subsubsection{Baseline systems}\label{sec_base}

 AMs in this series of experiments  are DNNs trained on BN features, and for the baseline with speaker adaptation  we used fMLLR adaptation. For these models, we also
 used two different training criteria: cross-entropy (CE) criterion and sequence-discriminative training with  Minimum Bayes Risk (sMBR) criterion in order to study the impact of the training criterion on the adaption performance. Hence, we trained four baseline DNN AMs (see~\ref{app1} for details):
 \begin{itemize}
\item \boldmath{$\mathrm{DNN}_\mathrm{BN}\textnormal{-}\mathrm{CE}$}:  BN features, CE criterion;
\item \boldmath{$\mathrm{DNN}_\mathrm{BN}\textnormal{-}\mathrm{sMBR}$}: 
BN features, sMBR criterion;
\item $\mathrm{DNN}_{\mathrm{BN\textnormal{-}fMLLR}}\textnormal{-}\mathrm{CE}$:
fMLLR-adapted BN features,  CE criterion;
\item $\mathrm{DNN}_{\mathrm{BN\textnormal{-}fMLLR}}\textnormal{-}\mathrm{sMBR}$:
fMLLR-adapted BN features,  sMBR criterion. 
\end{itemize}

 \textit{LM-cantab} was used for decoding.

\subsubsection{Proposed systems with speaker-adapted GMMD features}\label{sec_proposed_am_dnn}
For experiments with speaker adaptation, we trained 
four acoustic models using the approaches proposed in Sections~\ref{DNN-GMMD}.

\begin{itemize}
\item \boldmath{$\mathrm{DNN}_{\mathrm{GMMD}\oplus\mathrm{BN}}\textnormal{-}\mathrm{CE}$}: speaker-independent (SI) GMMD features appended with BNs, CE criterion; 
\item \boldmath{$\mathrm{DNN}_{\mathrm{GMMD}\oplus\mathrm{BN}}\textnormal{-}\mathrm{sMBR}$}:
SI GMMD features appended with BNs, sMBR criterion;
\item \boldmath{
$\mathrm{DNN}_{\mathrm{\mathrm{GMMD\textnormal{-}MAP}\oplus\mathrm{BN}}}\textnormal{-}\mathrm{CE}$}: SAT DNN model trained on speaker adapted GMMD-MAP features, CE criterion.
\item 
$\mathrm{DNN}_{\mathrm{\mathrm{GMMD\textnormal{-}MAP}\oplus\mathrm{BN}}}\textnormal{-}\mathrm{sMBR}$: on speaker adapted GMMD-MAP features, sMBR criterion.
\end{itemize}
\boldmath{}
Models
$\mathrm{DNN}_{\mathrm{\mathrm{GMMD\textnormal{-}MAP}\oplus\mathrm{BN}}}\textnormal{-}\mathrm{CE}$
and 
$\mathrm{DNN}_{\mathrm{GMMD\textnormal{-}MAP}\oplus\mathrm{BN}}\textnormal{-}\mathrm{sMBR}$
were trained as described in Section~\ref{DNN-GMMD}.
The GMMD features were extracted using a monophone auxiliary GMM model, trained on BN features. This GMM model was adapted for each speaker by MAP adaptation algorithm (Section~\ref{sec_map}). See more details in~\ref{app3}.

\unboldmath

\subsubsection{Adaptation and fusion results}\label{sec_dnn_results}
The adaptation experiments were conducted in an unsupervised mode on the test data using  transcripts from the first decoding pass obtained by the baseline SAT-DNN model, unless explicitly stated otherwise.

We empirically studied different types of fusion described in Section \ref{Sect_Fus} and applied them to 
DNN models trained using GMMD-features extracted as proposed in Section \ref{Sect_SAT}.
The performance results in terms of WER for SI and SAT DNN-HMM models are presented in Table~\ref{table_results_dnn_adapt}.
The first four lines of the table correspond to the baseline SI (\#1, \#2) and SAT (\#3, \#4) DNNs, which were trained as
described in Section \ref{sec_base}.

\begin{table}
\caption{ Summary of the adaptation results for DNN models. The results in parentheses correspond to WER of the consensus hypothesis.}\label{table_results_dnn_adapt}
\centerline{\renewcommand{\tabcolsep}{1.0mm}
		\begin{tabular}{|c|l|c|c|c|c|}
		\hline
		\multirow{2}{*}{\#} & \multirow{2}{*}{Features} & \multirow{2}{*}{DNN} &   \multicolumn{3}{|c|}{WER,\%} \\ \cline{4-6} 
		& & & Development & $\mathrm{Test_1}$ & $\mathrm{Test_2}$ \\  \hline 
        1 & BN & CE  & 	 13.16         & 11.94  & 15.43\\ 
		2 & BN & sMBR  & 	 12.14          & 10.77  & 13.75\\ \hline
        3 & BN-fMLLR & CE  & 	 11.72         & 10.88  & 14.21\\ 
        4 & BN-fMLLR & sMBR    & 	 10.64 (10.57)  & 9.52 (9.46)  & 12.78 (12.67)\\  \hline \hline
        5 & GMMD$\oplus$BN & CE    & 	 12.92          & 11.62 & 15.19\\
        6 & GMMD$\oplus$BN & sMBR    & 	 11.80          & 10.47  & 13.52\\ \hline
        7 & GMMD-MAP$\oplus$BN & CE     & 	 10.46    & 9.74  & 13.03\\ 
        8 & GMMD-MAP$\oplus$BN & sMBR    &  \textbf{10.26 (10.23)}  &\textbf{ 9.40 (9.31)}  & \textbf{12.52 (12.46)}\\ \hline
\end{tabular}}
\end{table}

Parameter $\tau$  in MAP adaptation, that controls the balance between the maximum likelihood estimate of the mean and its prior value~\cite{gauvain1994maximum, tomashenko2014speaker}, for both acoustic model training  and decoding was set equal to $5$. 

For comparison purpose with lattice-based fusion we report WER of the consensus hypothesis in parentheses for experiments \#4 and \#8.

After that we made posterior fusion of the obtained  model \#8 and the baseline SAT-DNN model (\#4). 
The result is given in Table~\ref{table_results_dnn_fus}, line \#9.
Value $\alpha$ in Formula~(\ref{eq_fus}) (Section \ref{Sect_Fus}) is a weight of the baseline SAT-DNN model. Parameter $\alpha$ was optimized on the development set.

\begin{table}
\caption{ Summary of the fusion results for DNN models. The results in parentheses correspond to WER of the consensus hypothesis. Here  $\downarrow$ denotes relative WER reduction (for consensus hypothesis) in comparison with AM trained on BN-fMLLR (\#4 in Table~\ref{table_results_dnn_adapt}). The bold figures in the table indicate the best performance improvement.}\label{table_results_dnn_fus}
\centerline{
		\begin{tabular}{|c|c|c|c|c|}
		\hline
		\multirow{2}{*}{\#}  & \multirow{2}{*}{\shortstack{ Fusion: \\ \\ \#4 and \#8}} &   \multicolumn{3}{|c|}{WER,\%} \\ \cline{3-5} 
		 &  & Development & $\mathrm{Test_1}$ & $\mathrm{Test_2}$ \\  \hline 
       9 & \shortstack{ \\ Posterior fusion,\\ $\alpha=0.45$ } & \shortstack{ \\ 9.98 (9.91) \\ $\downarrow$ \textbf{6.2} } & \shortstack{ \\ 9.15 (9.06) \\ $\downarrow \textbf{4.3}$ } & \shortstack{ \\ 12.11 (12.04) \\ $\downarrow$ \textbf{5.0} }\\ \hline
       10 & \shortstack{ \\ Lattice fusion, \\ $\alpha=0.46$}  &  \shortstack{ \\ 10.06 \\ $\downarrow 4.8$  } & \shortstack{ \\ 9.09 \\ $\downarrow 4.0$ } & \shortstack{ \\ 12.12 \\ $\downarrow 4.4$ } \\ \hline
\end{tabular}}
\end{table}

Finally, we applied lattice fusion for the same pair of models (line \#10). 
In this type of fusion, before merging lattices, for each edge, scores were replaced by its a posteriori probabilities. Posteriors were computed for each lattice independently. The optimal normalizing factors for each model were found independently on
the development set. Then the two lattices were merged into a single lattice and posteriors were weighted using parameter $\alpha$. 
As before, value $\alpha$ in Formula (\ref{eq_fus}) corresponds to the baseline SAT-DNN model. 
The resulting lattice was converted into the CN and the final result was obtained from this CN. 

We can see, that both - posterior and lattice types of fusion provide similar improvement for all three models:
approximately 4\%--6\% of relative WER reduction (WERR) in comparison with the adapted baseline model (SAT DNN on fMLLR features, \#4), and 
12\%--18\% 
of relative WERR
in comparison with the SI baseline model (\#2).
For models \#7--8 only MAP adaptation was applied. Experiments \#9--10 present combination of two different adaptation types: MAP and fMLLR. 
Is is interesting to note that in all experiments optimal value of $\alpha$ is close to 0.5, so all types of models are 
equally important for fusion.
We can see that MAP adaptation on GMMD features can be complementary to fMLLR adaptation on conventional BN features.

\subsection{Experiments with TDNN models}\label{sec:experim_tdnn}

In this section, we  expand our experimental study to the TDNN topology.

\subsubsection{Baseline system}

We trained four baseline TDNN acoustic models, which differ only in the type of the input features (see~\ref{app2} for details):
\begin{itemize}
\item \boldmath{$\mathrm{TDNN}_\mathrm{MFCC}$}: 
 high-resolution MFCC features;
\item \boldmath{$\mathrm{TDNN}_{\mathrm{MFCC}\oplus\mathrm{i}\textnormal{-}\mathrm{vectors}}$}:
 high-resolution MFCC features appended with 100-dimensional i-vectors;
\item \boldmath{$\mathrm{TDNN}_{\mathrm{BN}}$}: BN features;
\item \boldmath{$\mathrm{TDNN}_{\mathrm{BN}\oplus\mathrm{i}\textnormal{-}\mathrm{vectors}}$}: BN features appended with 100-dimensional i-vectors.
\end{itemize}

\textit{LM-lium} was used for decoding.

\subsubsection{Proposed systems with speaker-adapted GMMD features}\label{sec_proposed_am_tdnn}

Four TDNNs were tarined using GMMD features as proposed in Section~\ref{TDNN-GMMD}:
\begin{itemize}
\item \boldmath{$\mathrm{TDNN}_{\mathrm{MFCC}\oplus\mathrm{GMMD}}$}:  high-resolution MFCC features appended with speaker adapted GMMD features;
\item \boldmath{$\mathrm{TDNN}_{\mathrm{MFCC}\oplus\mathrm{GMMD}\oplus\mathrm{i}\textnormal{-}\mathrm{vectors}}$}: high-resolution MFCC features appended with speaker adapted GMMD features and  100-dimensional i-vectors;
\item \boldmath{$\mathrm{TDNN}_{\mathrm{BN}\oplus\mathrm{GMMD}}$}: BN features appended with speaker adapted GMMD features;
\item \boldmath{$\mathrm{TDNN}_{\mathrm{BN}\oplus\mathrm{i}\textnormal{-}\mathrm{vectors}\oplus\mathrm{GMMD}}$} is a version of the  \boldmath{$\mathrm{TDNN}_{\mathrm{BN}\oplus\mathrm{GMMD}}$} with 100-dimensional i-vectors.
\end{itemize}
All the four TDNN models were trained in the same manner, as the baseline TDNN model, and differ only in the type of the input features.

\subsubsection{Results for TDNN models}\label{sec_tdnn_results}

\begin{table}
\caption{ Summary of the adaptation results for TDNN models. The results in parentheses correspond to WER of the consensus hypothesis. The bold figures in the table indicate the best performance improvement.}\label{table_results_tdnn_adapt}
\centerline{
\setlength{\tabcolsep}{4pt}
		\begin{tabular}{|c|l|c|c|c|c|}
		\hline
		\multirow{2}{*}{\#} & \multirow{2}{*}{Features} &   \multicolumn{3}{|c|}{WER,\%} \\ \cline{3-5} 
		& &  Development & $\mathrm{Test_1}$ & $\mathrm{Test_2}$ \\  \hline 
        1 & MFCC                                      & 	 11.98 (11.88)   & 9.42 (9.31) & 12.77 (12.66)\\ 
		2 & MFCC$\oplus$i-vectors                   & 	 10.16 (10.12)   & 7.98 (7.90) & 11.73 (11.70)\\  \hline
        3 & BN                                        & 	 11.68 (10.62)   & 8.83 (8.76) & 12.44 (12.41) \\ 
        4 & BN$\oplus$i-vectors                     & 	 10.05 (9.98)    & 8.29 (8.21) & 11.90 (11.88)\\  \hline  \hline
        5 & GMMD$\oplus$MFCC                       & 	 9.79 (9.70)     & 8.26 (8.17) & 12.21 (12.16)\\
        6 & GMMD$\oplus$MFCC$\oplus$i-vectors   & 	 9.35 (9.32)     & 8.10 (8.05) & 11.98 (11.95) \\ \hline
        7 & GMMD$\oplus$BN                         & 	 9.57 (9.52)     & 8.18 (8.13) & 11.92 (11.87) \\
        8 & GMMD$\oplus$BN$\oplus$i-vectors     & 	\textbf{ 9.41 (9.34)}     & \textbf{7.85 (7.74)} & \textbf{11.70 (11.65)}\\ \hline
\end{tabular}}
\end{table}
In this set of experiments, for adaptation of the proposed models without i-vectors, we used the first decoding output made by the baseline model, which is also without i-vectors (\boldmath{$\mathrm{TDNN}_{\mathrm{BN}}$}).

Adaptation results for the TDNN models are given in Table~\ref{table_results_tdnn_adapt}.
The best result was obtained by 
\boldmath{$\mathrm{TDNN}_{\mathrm{GMMD}\oplus\mathrm{BN}\oplus\mathrm{i}\textnormal{-}\mathrm{vectors}}$} 
(line \#8).
\unboldmath{}
For Development set it gives $7.4\%$ of relative WERR over \boldmath{$\mathrm{TDNN}_{\mathrm{MFCC}\oplus\mathrm{i}\textnormal{-}\mathrm{vectors}}$}, though for the other test sets the result is very close to the baseline (line \#2) for the models which are already MAP-adapted. \unboldmath{}
If we compare (\#5, \#6) or (\#7, \#8), we can see that i-vectors always give an additional improvement for the model which is already MAP-adapted.

To further investigate the complementary of the two different adaptation techniques,  we performed CNC of recognition results for different TDNN models (Table~\ref{table_results_tdnn_fus}).
The best result, obtained by combination of TDNN models \#2 and \#8, provides approximately $7$-$13\%$ of relative WERR in comparison with the baseline model \boldmath{$\mathrm{TDNN}_{\mathrm{MFCC}\oplus\mathrm{i}\textnormal{-}\mathrm{vectors}}$}. \unboldmath{}

\begin{table}
\caption{Summary of the fusion  results (CNC) for TDNN models. Here  $\downarrow$ denotes relative WER reduction (for consensus hypothesis) in comparison with the baseline \boldmath{$\mathrm{TDNN}_{\mathrm{MFCC}\oplus\mathrm{i}\textnormal{-}\mathrm{vectors}}$},
\unboldmath
$\alpha$ is a weight of $\textnormal{TDNN}_{1}$ in the fusion. The bold figures in the table indicate the best performance improvement.}\label{table_results_tdnn_fus}
\centerline{
		\begin{tabular}{|c|c|c|c|c|c|c|}
		\hline
		\multirow{2}{*}{\#}  & \multirow{2}{*}{$\textrm{TDNN}_{1}$} &  \multirow{2}{*}{$\textrm{TDNN}_{2}$} & \multirow{2}{*}{$\alpha$}  & \multicolumn{3}{|c|}{WER,\%} \\ \cline{5-7} 
	&	& &  & Development & $\mathrm{Test_1}$ & $\mathrm{Test_2}$ \\  \hline 
9  & 2 & 4	 &	0.50  &  	9.34 $\downarrow 7.7$ &		7.52 $\downarrow 4.8$ & 11.20 $\downarrow 4.2$ \\ 
10 & 4 & 8	 & 	0.31  &  	9.29 $\downarrow 8.2$ &		7.68 $\downarrow 2.8$ & 11.32 $\downarrow 3.3$ \\
11 & 8 & 6	 & 	0.33  & 	9.28 $\downarrow 8.3$ & 	7.81 $\downarrow 1.2$ & 11.54 $\downarrow 1.4$ \\
12 & 4 & 6   & 	0.38  &  	9,22 $\downarrow 8.9$ & 	7.85 $\downarrow 0.7$ & 11.39 $\downarrow 2.7$ \\
13 & 2 & 7	 & 	0.49  &  	8.96 $\downarrow 11.5$ & 	7.33 $\downarrow 7.3$ & 10.95 $\downarrow 6.4$ \\
14 & 2 & 6	 & 	0.50  &  	8.92 $\downarrow 11.8$ &   	7.33 $\downarrow 7.3$ & 10.99 $\downarrow 6.1$ \\
15 & 2 & 8	 & 	0.46  &  	\textbf{8.84} $\downarrow$ \textbf{12.7}  &    \textbf{7.28} $\downarrow$ \textbf{7.9} & \textbf{10.91} $\downarrow$ \textbf{6.8} \\ \hline 
\end{tabular}}
\end{table}

\section{Feature analysis}\label{sec_feature_analysis}
The objective of this section is to analyze the proposed GMMD features and the adaptation algorithm for better understanding their nature and properties at different levels.

\subsection{Lattice-based features}\label{sec_lattice-based-features}
Lattice based features or time dependent state posterior scores~\cite{uebel2001improvements, gollan2008confidence} are obtained from computing arc posteriors from the output lattices of the decoder.
These features contain more information about the decoding process than the posterior probabilities from neural networks because in their extraction also language model probabilities, likelihoods of decoding hypothesis, and other information are taken into account. 
We use this type of features to analyze the quality of the adaptation of TDNN acoustic models.

Let $\left\{ph_{1},\ldots,ph_{M}\right\}$ be a set of phonemes and the silence model. 
For each time frame $t$, we calculate $p_t^m$~--- the
confidence score of phoneme $ph_m$ ($1\leq m \leq M$) at time $t$
 in the decoding lattice 
by calculating arc posterior probabilities. 
The forward-backward algorithm is used to calculate these arc posterior probabilities from the lattice as follows:
\begin{equation}
p(l|O)=\frac{\sum_{q\in Q_{l}}{p_{ac}(O|q)^{\frac{1}{\lambda}}p_{lm}(w)}}{p(O)},
\label{eq_poster}
\end{equation}
where $\lambda$ is the 
scale factor (the optimal value of $\lambda$ is found empirically by minimizing WER of the consensus hypothesis \cite{mangu2000finding});
$q$ is a path through the lattice corresponding to the word sequence $w$;
$Q_{l}$ is the set of paths passing through arc~$l$;
$p_{ac}(O|q)$ is the acoustic likelihood;
$p_{lm}(w)$ is the language model probability;
and $p(O)$ is the overall likelihood of all paths through the lattice.

For the given frame $\textbf{o}_t$ at time $t$, we calculate its probability $p(\textbf{o}_{t})\in ph_{m}$ of belonging to phoneme  $ph_m$, using  lattices obtained  from the first decoding pass:

\begin{equation}
p_t^m=p(\textbf{o}_{t}\in ph_{m})=\sum_{l\in S_{m}(\textbf{o}_{t})}{p(l|O)},
\label{eq_prob_for_weight}
\end{equation}
where $S_{m}(\textbf{o}_{t})$ is the set of all arcs corresponding to the phoneme $ph_m$ in the lattice at time $t$;
$p(l|O)$ is the posterior probability of arc $l$ in the lattice.

The obtained probability $p(\textbf{o}_{t}\in ph_{m})$ of frame $\textbf{o}_t$ belonging to phoneme $ph_m$ is the component value $p_t^m$ on the new feature vector $\textbf{p}_t$. 
Thus for a given acoustic feature vector $\textbf{o}_{t}$ at time $t$ we obtain a new feature vector $\textbf{p}_t$:
\begin{equation}\label{eq:p_t}
\textbf{p}_t=\left(p_t^{1},\ldots,p_t^M\right),
\end{equation}
where $M$ is the number of phones in the phoneme set used in the ASR system.

Hence for each frame $\textbf{o}_{t}$ we have a $M$-dimensional vector~$\textbf{p}_{t}$, each component of which represents the probability of this frame to belong to a certain
phoneme.
When some phonemes are not present in the lattice for a certain frame, we set 
 probabilities equal to some very small value  $\epsilon$ for them in the vector (where $\epsilon$ is a minimum value from lattices: $\epsilon\approx 10^{-9}$). 
 In addition to this, we use state index information (position of the state in phoneme HMM: 0, 1 or 2)
from the Viterbi alignment from the original transcripts.

\subsection{Davies-Bouldin index}\label{sec_db_inex}
We use the Davies-Bouldin (DB) index~\cite{davies1979acluster} to evaluate the quality of the phoneme state clusters obtained from the latticed-based features.

\begin{equation}
DB=\frac{1}{K}\sum_{k=1}^{K}\max_{j\neq k}\left(\frac{\sigma_k+\sigma_j}{\rho_{k,j}}\right),
\end{equation}
where

$K$ is the number of clusters;

$\sigma_k$ is the scatter within the cluster $k$,  which is our case the standard deviation of
the distance of all
vectors
corresponding to cluster $k$, to the cluster center (other possible metric variants are described in~\cite{davies1979acluster});

$\rho_{k,j}$ is a between-cluster separation measure, which in our case is the Euclidean distance between the centroids of clusters $k$ and $j$.

Smaller values of DB index correspond to better clusters.

\subsection{Visual analysis using t-SNE}\label{sec_tsne}
The lattice-based features were visualized using t-distributed stochastic neighbor embedding  (t-SNE) analysis~\cite{maaten2008visualizing}. 
This technique allows us to visualize high-dimensional data into two or three dimensional space, in such a way that the vectors, which are close in the original space, are  also close in the low dimensional t-SNE representation. 

We are interested in how well the different acoustic models can cluster different phoneme states.
For better visualization, we used data only from inter-word phones 
and only from the middle state of HMM (State $1$).
We choose only those phonemes for which we have sufficient amount of data for analysis and perform t-SNE analysis independently on three different groups of phonemes\footnote{The notations are given according to the ARPAbet phoneme set: https://en.wikipedia.org/wiki/Arpabet}:
\begin{itemize}
\item \it{Vowels}
(UH, OW, AO, EY, ER, AA, AY, IY, EH, AE, IH, AH);
\item \it{Consonants-1}: 
Liquids (L, R), 
Nasals (M, N, NG), 
Semivowels (W);
\item \it{Consonants-2}: 
Stops (P, T, D, K),
Affricates (CH), 
Fricatives (F, V, TH, S, Z, SH). 
\end{itemize}

\subsection{Analysis for TDNN models}
\label{sec_anal_tdnn}
In this set of experiments we compare the following three TDNN acoustic models: 
\boldmath{$\mathrm{TDNN}_\mathrm{MFCC}$},
\boldmath{$\mathrm{TDNN}_{\mathrm{BN}\oplus\mathrm{i}\textnormal{-}\mathrm{vectors}}$}
and
\boldmath{$\mathrm{TDNN}_{\mathrm{BN}\oplus\mathrm{i}\textnormal{-}\mathrm{vectors}\oplus\mathrm{GMMD}}$}.
\unboldmath{}
All the experiments described in this section (except for Figures~\ref{fig_byspeakers} and \ref{fig_depend_on_wer}) are performed using lattice-based features (Section~\ref{sec_lattice-based-features}) on the \textit{Development} data set.

\begin{figure}[h]
\begin{center}
\begin{subfigure}{0.046\textwidth}
\center{\includegraphics[width=\textwidth, trim = 135mm 78mm 18mm 80mm]{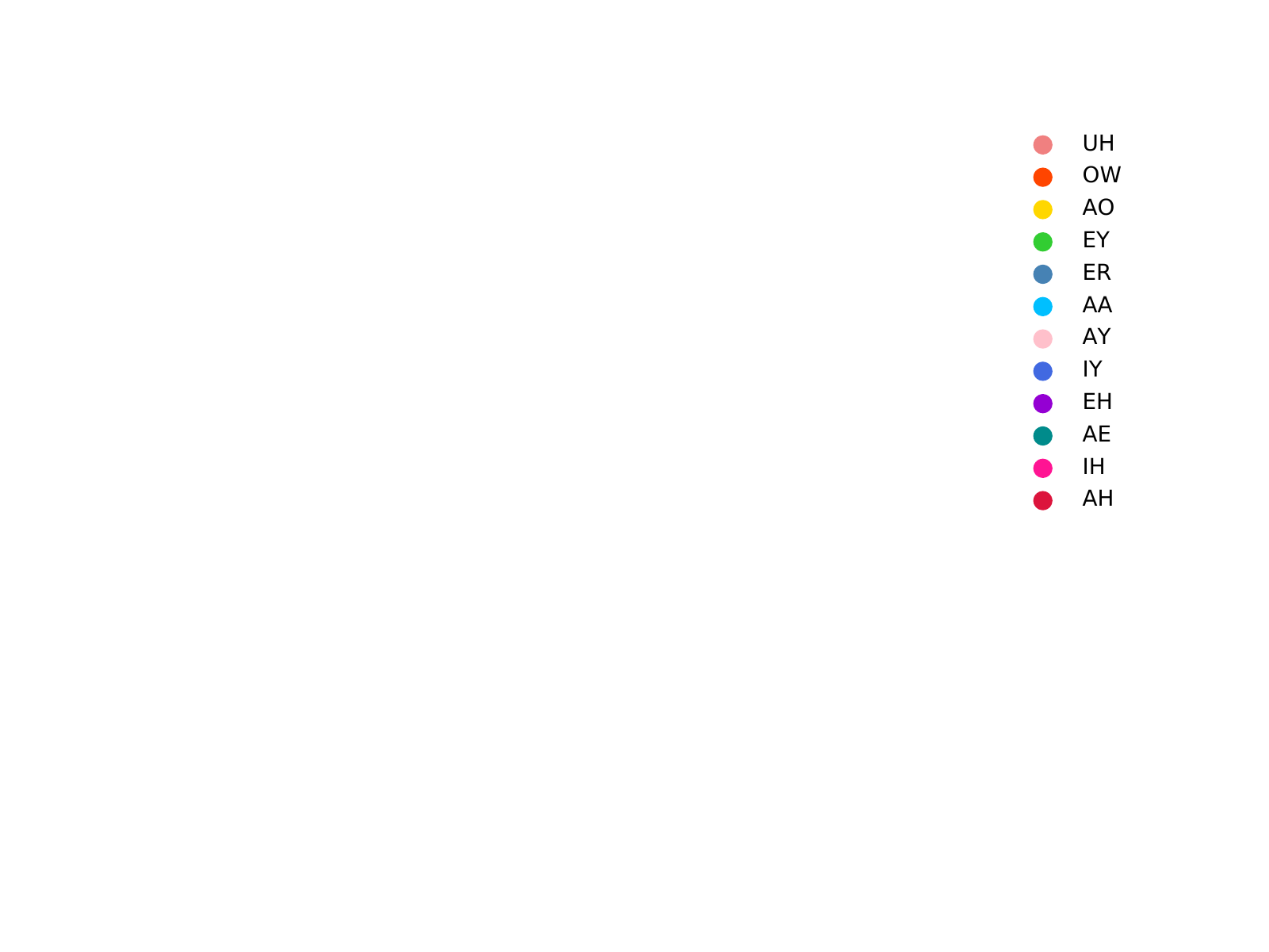}}
\end{subfigure}
\begin{subfigure}{0.296\textwidth}
\center{\includegraphics[width=\textwidth, trim = 20mm 18mm 18mm 20mm,  clip]{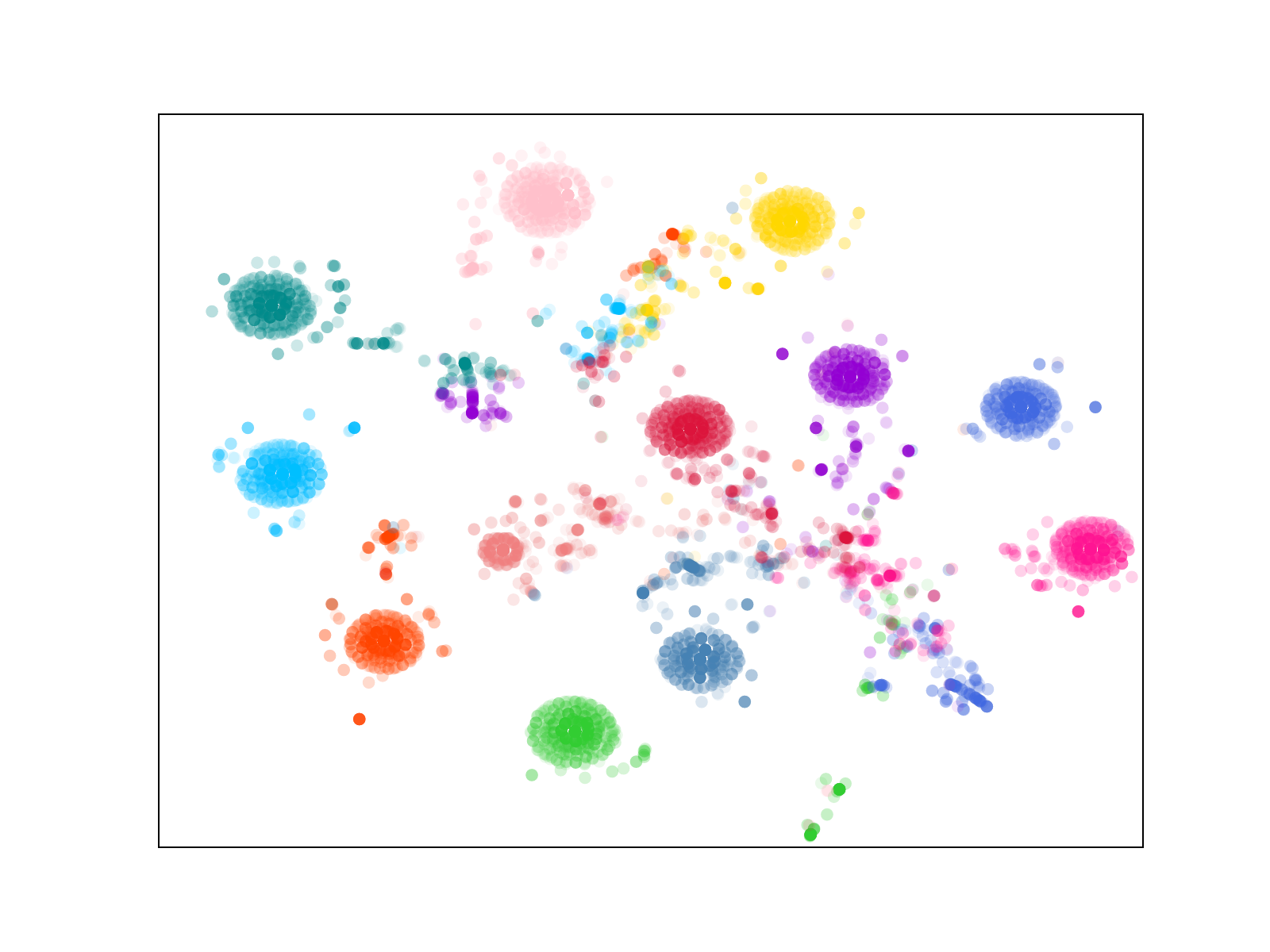}}
\caption{$\mathrm{MFCC}$}
\label{mfcc_vowels}
\end{subfigure}
\begin{subfigure}{0.296\textwidth}
\center{\includegraphics[width=\textwidth, trim = 20mm 18mm 20mm 20mm, clip]{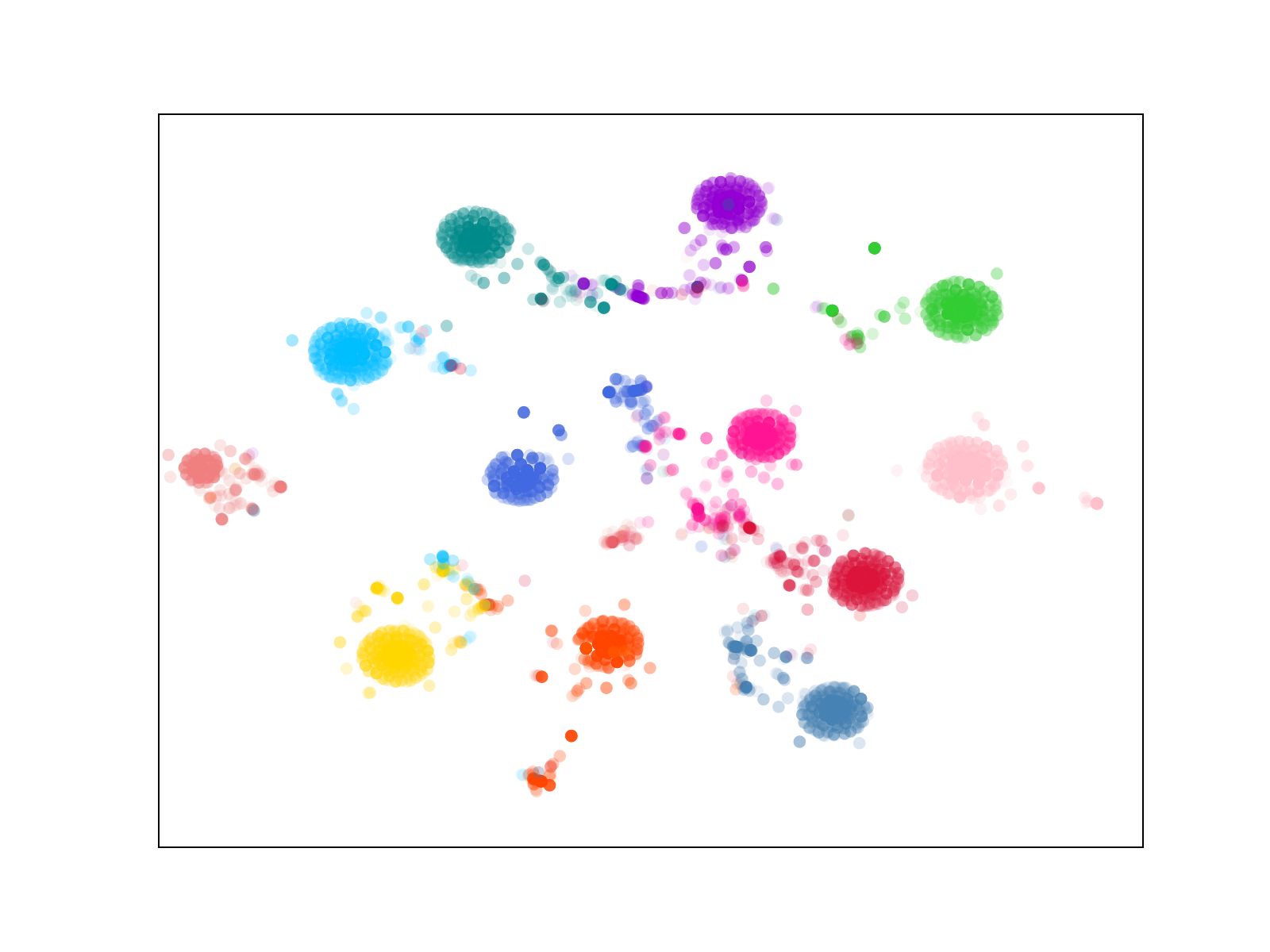}}
\caption{$\mathrm{BN}\oplus\mathrm{i}\textnormal{-}\mathrm{vectors}$}
\label{bn_ivect_vowels}
\end{subfigure}
\begin{subfigure}{0.295\textwidth}
\center{\includegraphics[width=\textwidth, trim =  18mm 18mm 20mm 20mm, clip]{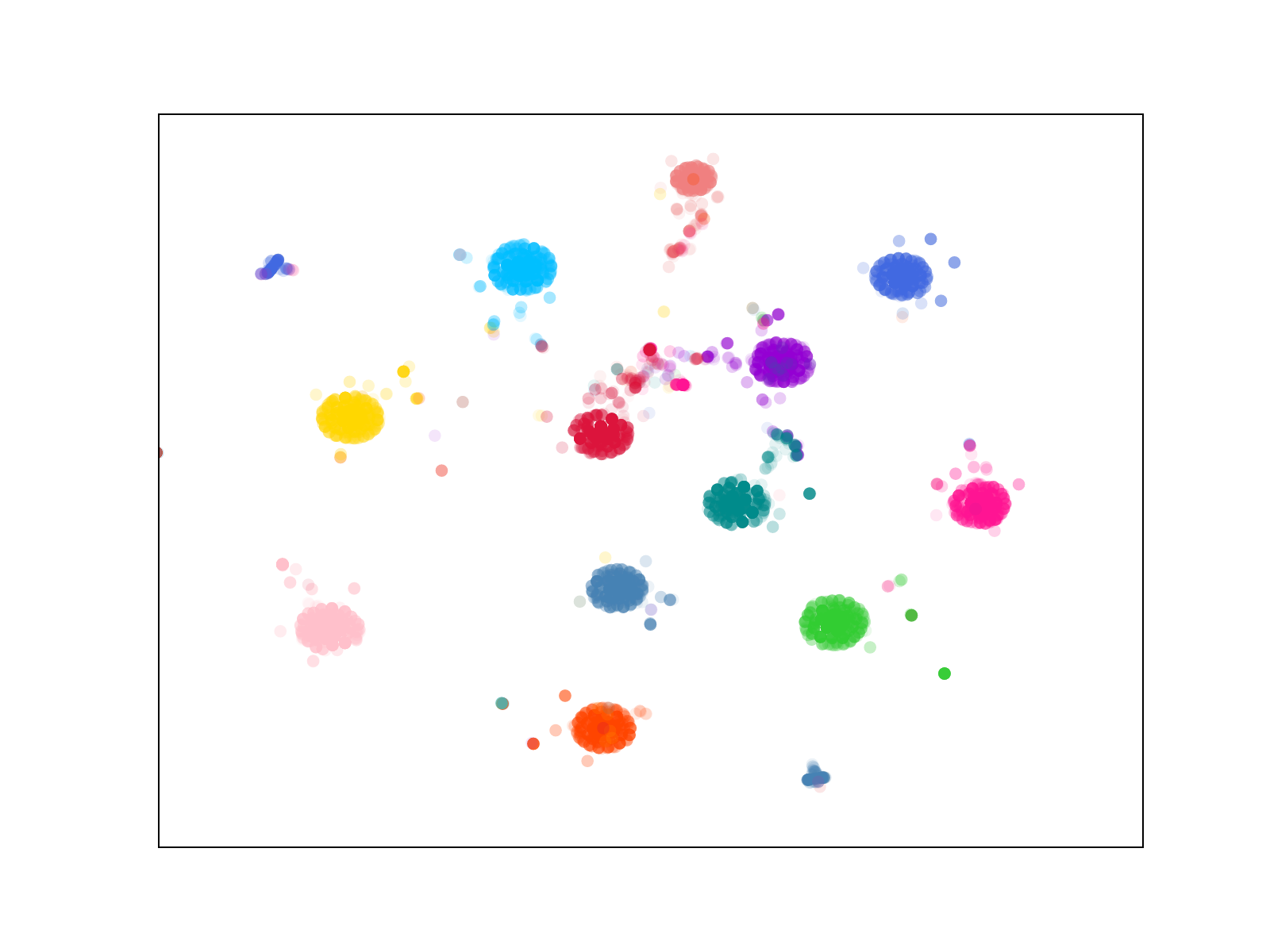}}
\caption{$\mathrm{BN}\oplus\mathrm{i}\textnormal{-}\mathrm{vect}\oplus\mathrm{GMMD}$}
\label{gmmd_bn_ivect_vowels}
\end{subfigure}
\end{center}
\caption{Analysis of lattice-based features for \textit{\textbf{vowels}} using t-SNE for TDNN models trained on different basic features.}
\label{fig_tsnev}
\end{figure}

\begin{figure}[h]
\begin{center}
\begin{subfigure}{0.046\textwidth}
\center{\includegraphics[width=\textwidth, trim = 135mm 78mm 18mm 80mm]{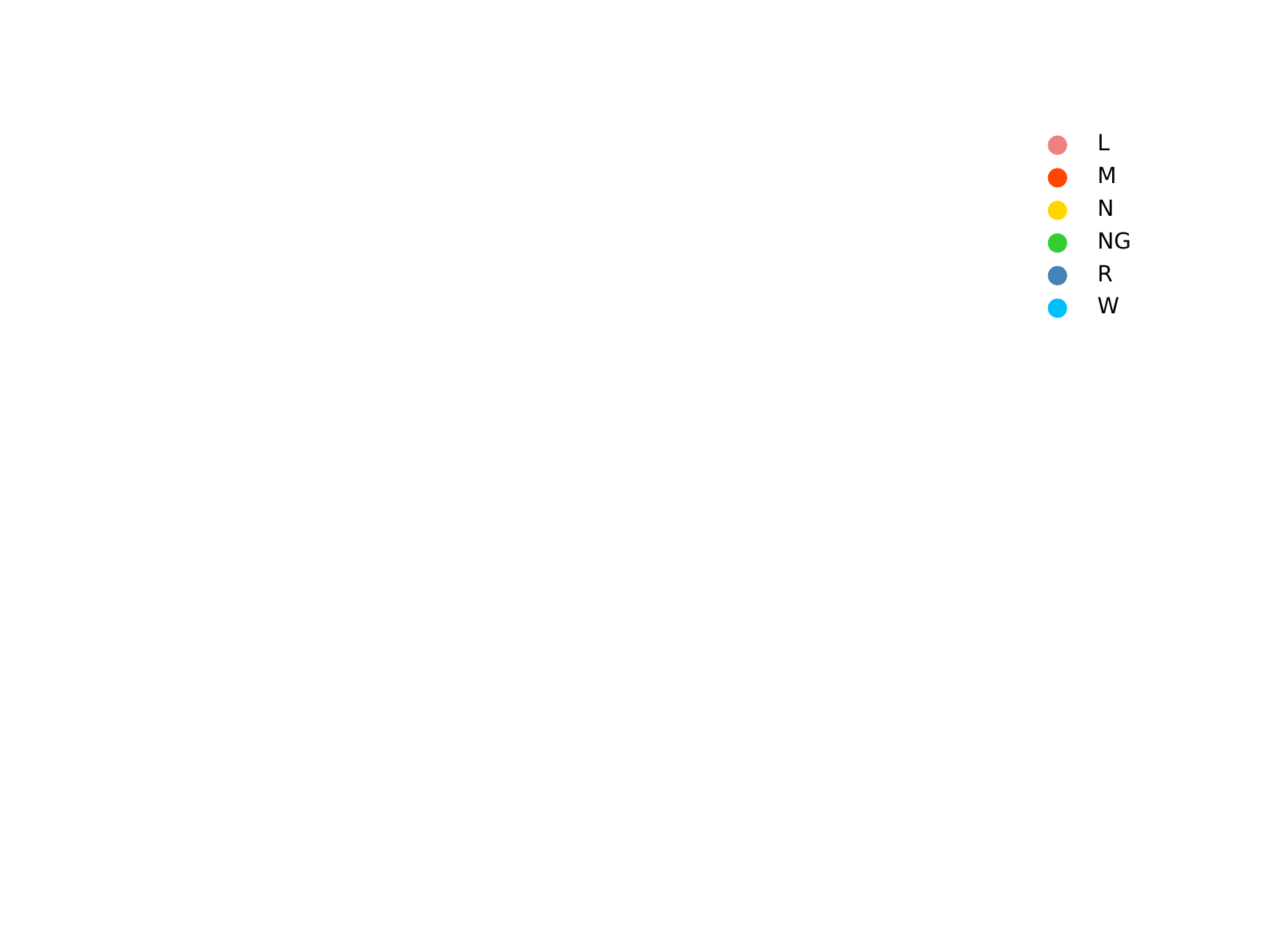}}
\end{subfigure}
\begin{subfigure}{0.3\textwidth}
\center{\includegraphics[width=\textwidth, trim =  18mm 18mm 20mm 20mm, clip]{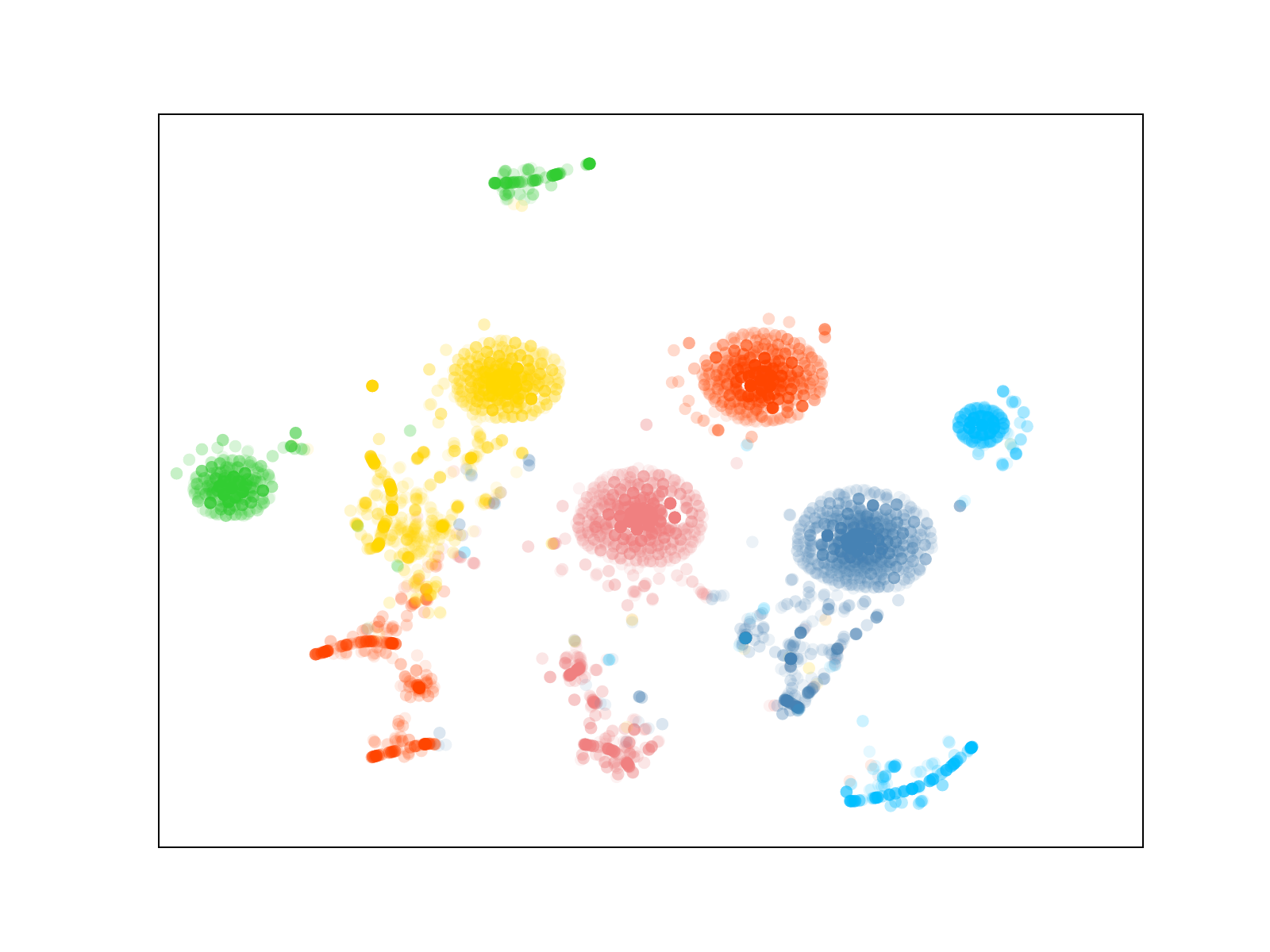}}
\caption{$\mathrm{MFCC}$}
\label{mfcc_cons1}
\end{subfigure}
\begin{subfigure}{0.3\textwidth}
\center{\includegraphics[width=\textwidth, trim =  18mm 18mm 20mm 20mm, clip]{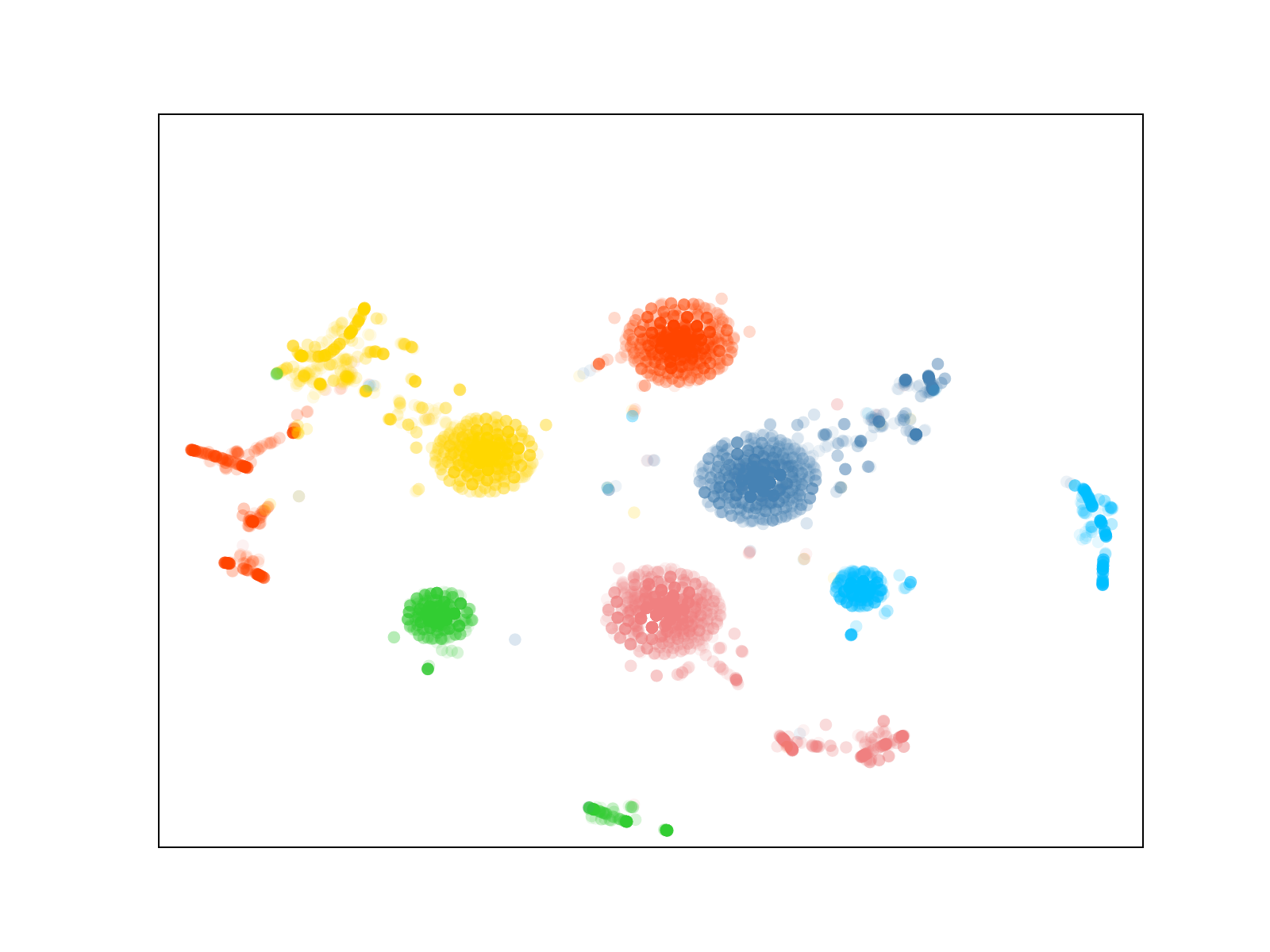}}
\caption{$\mathrm{BN}\oplus\mathrm{i}\textnormal{-}\mathrm{vectors}$}
\label{bn_ivect_cons1}
\end{subfigure}
\begin{subfigure}{0.3\textwidth}
\center{\includegraphics[width=\textwidth, trim =  18mm 18mm 20mm 20mm, clip]{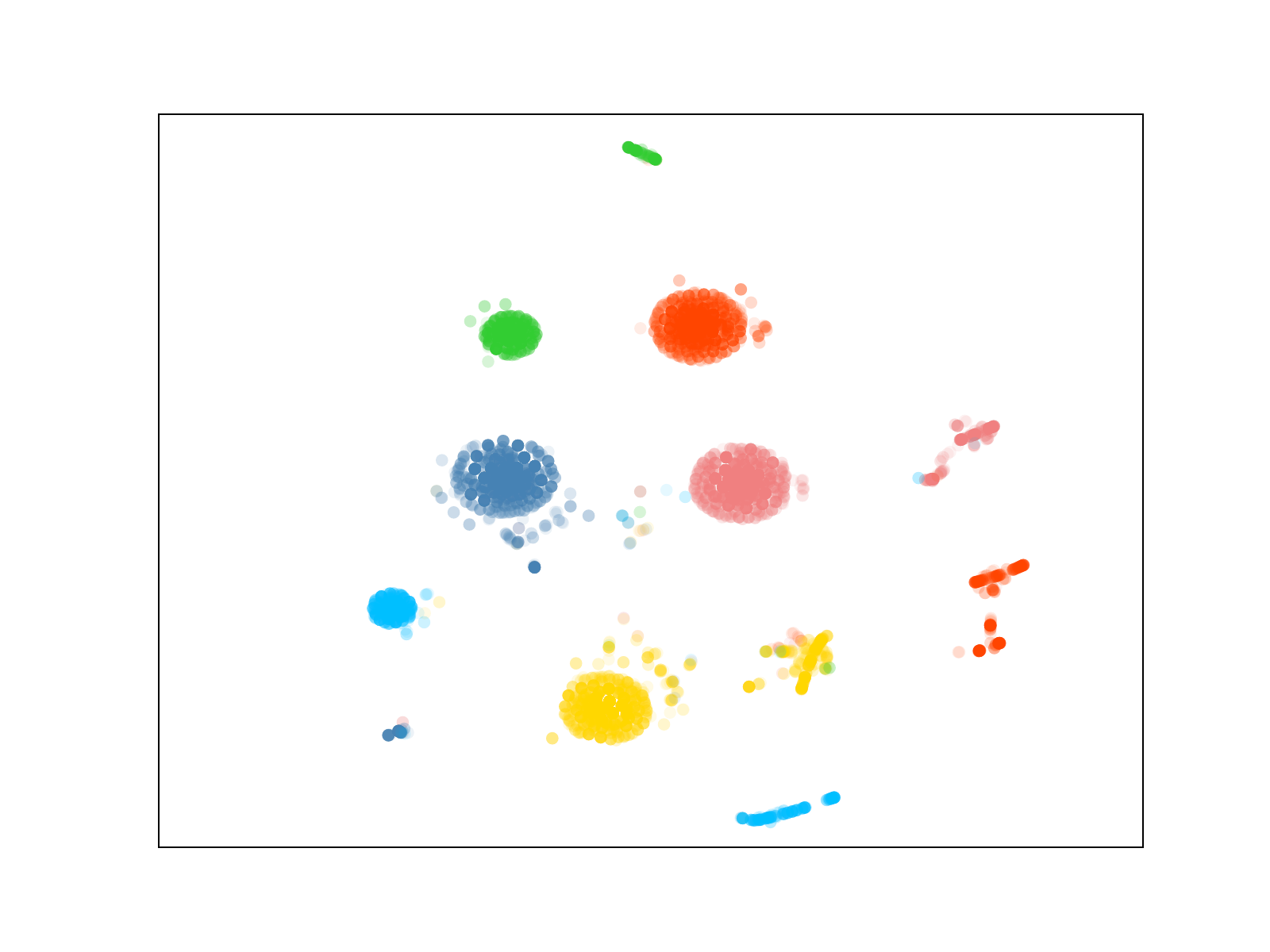}}
\caption{$\mathrm{BN}\oplus\mathrm{i}\textnormal{-}\mathrm{vect}\oplus\mathrm{GMMD}$}
\label{gmmd_bn_ivect_cons1}
\end{subfigure}
\end{center}
\caption{Analysis of lattice-based features for \textit{\textbf{consonants-1}} using t-SNE for TDNN models trained on different basic features.}
\label{fig_tsne_c1}
\end{figure}

\begin{figure}[h]
\begin{center}
\begin{subfigure}{0.05\textwidth}
\center{\includegraphics[width=1.1\textwidth, trim = 1mm 0mm 1mm 5mm]{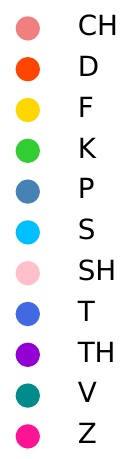}}
\end{subfigure}
\begin{subfigure}{0.3\textwidth}
\center{\includegraphics[width=\textwidth, trim =  18mm 18mm 20mm 20mm, clip]{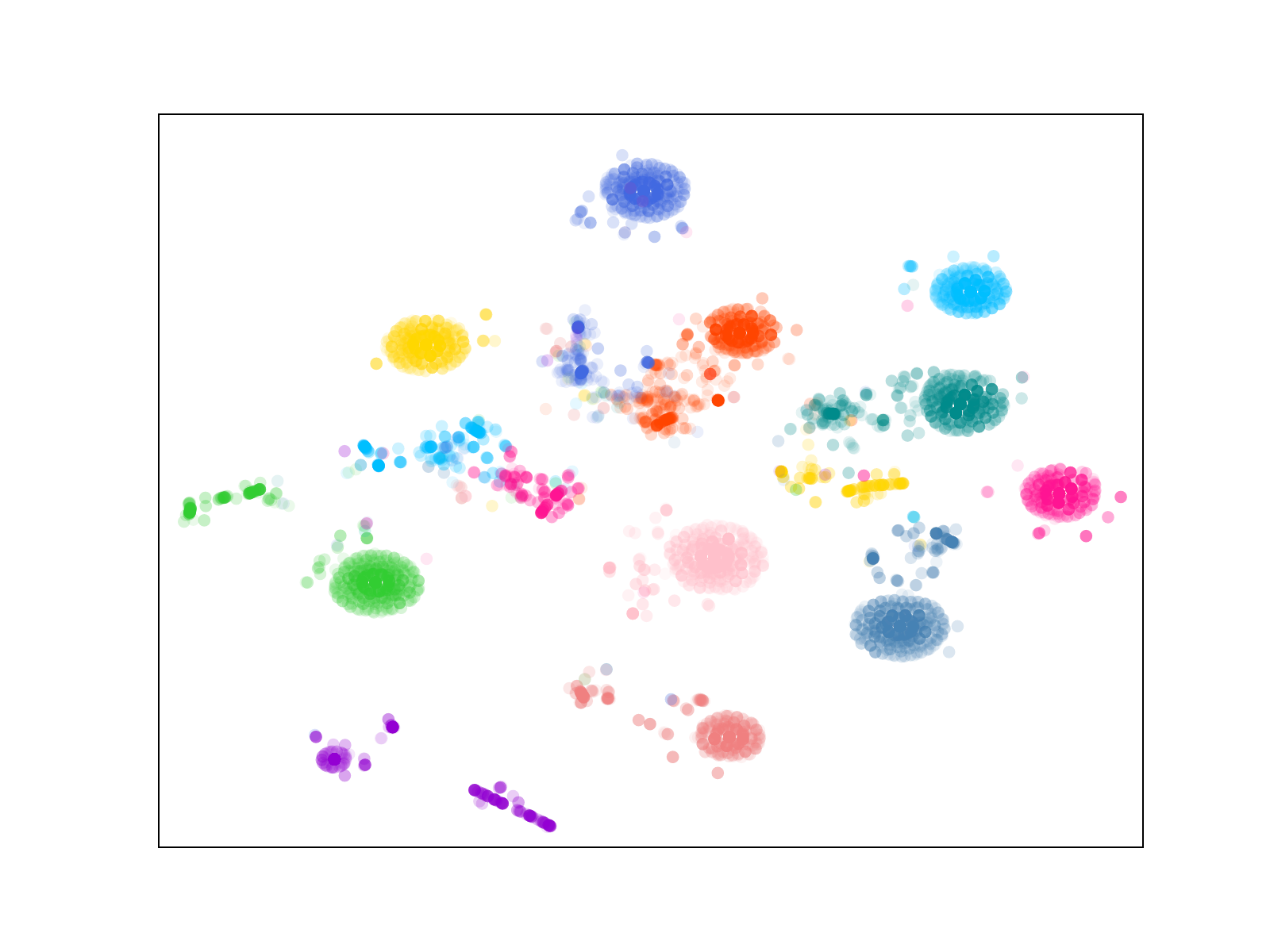}}
\caption{$\mathrm{MFCC}$}
\label{mfcc_cons2}
\end{subfigure}
\begin{subfigure}{0.3\textwidth}
\center{\includegraphics[width=\textwidth, trim =  18mm 18mm 20mm 20mm, clip]{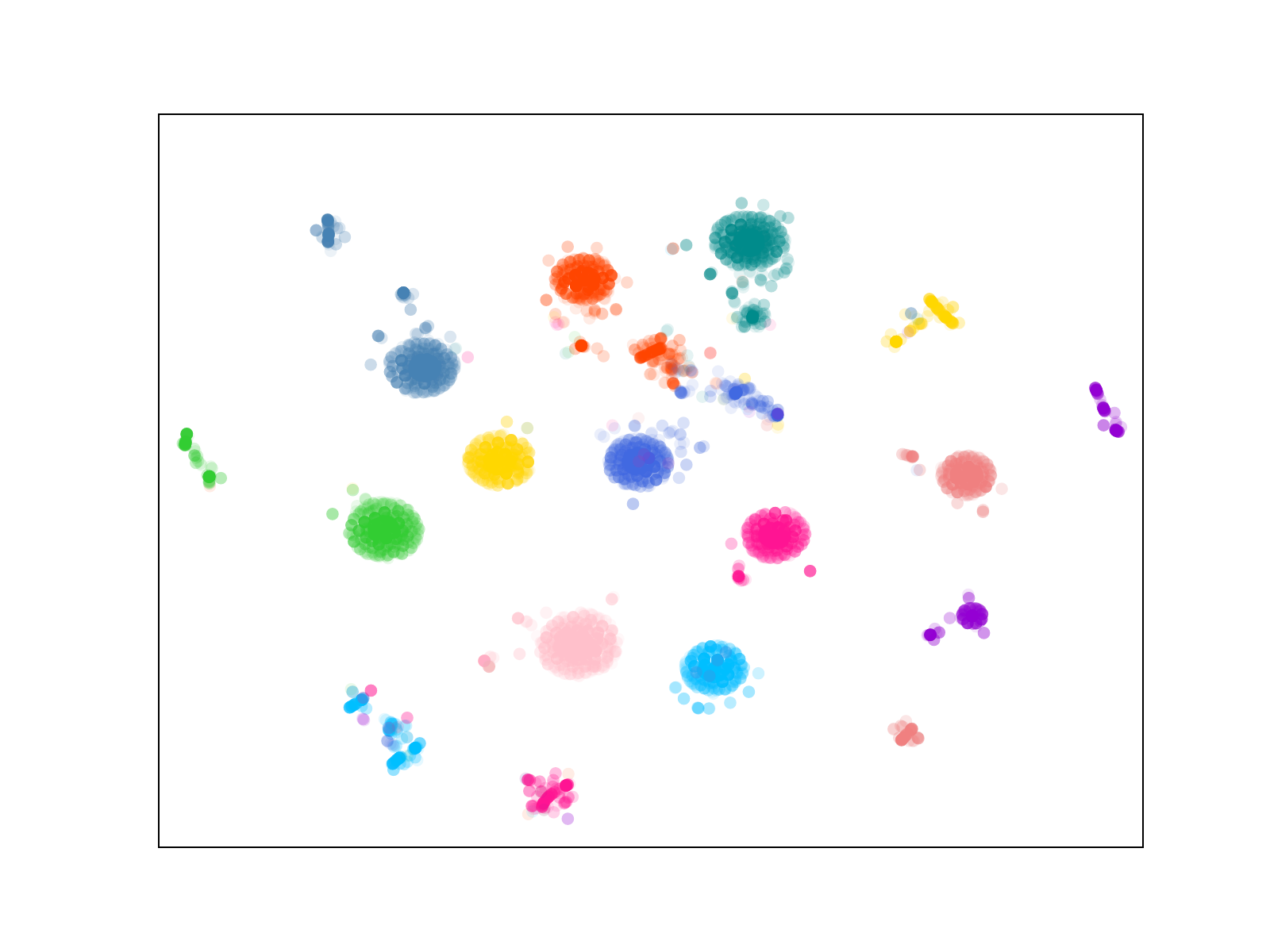}}
\caption{$\mathrm{BN}\oplus\mathrm{i}\textnormal{-}\mathrm{vectors}$}
\label{bn_ivect_cons2}
\end{subfigure}
\begin{subfigure}{0.3\textwidth}
\center{\includegraphics[width=\textwidth, trim =  18mm 18mm 18mm 20mm, clip]{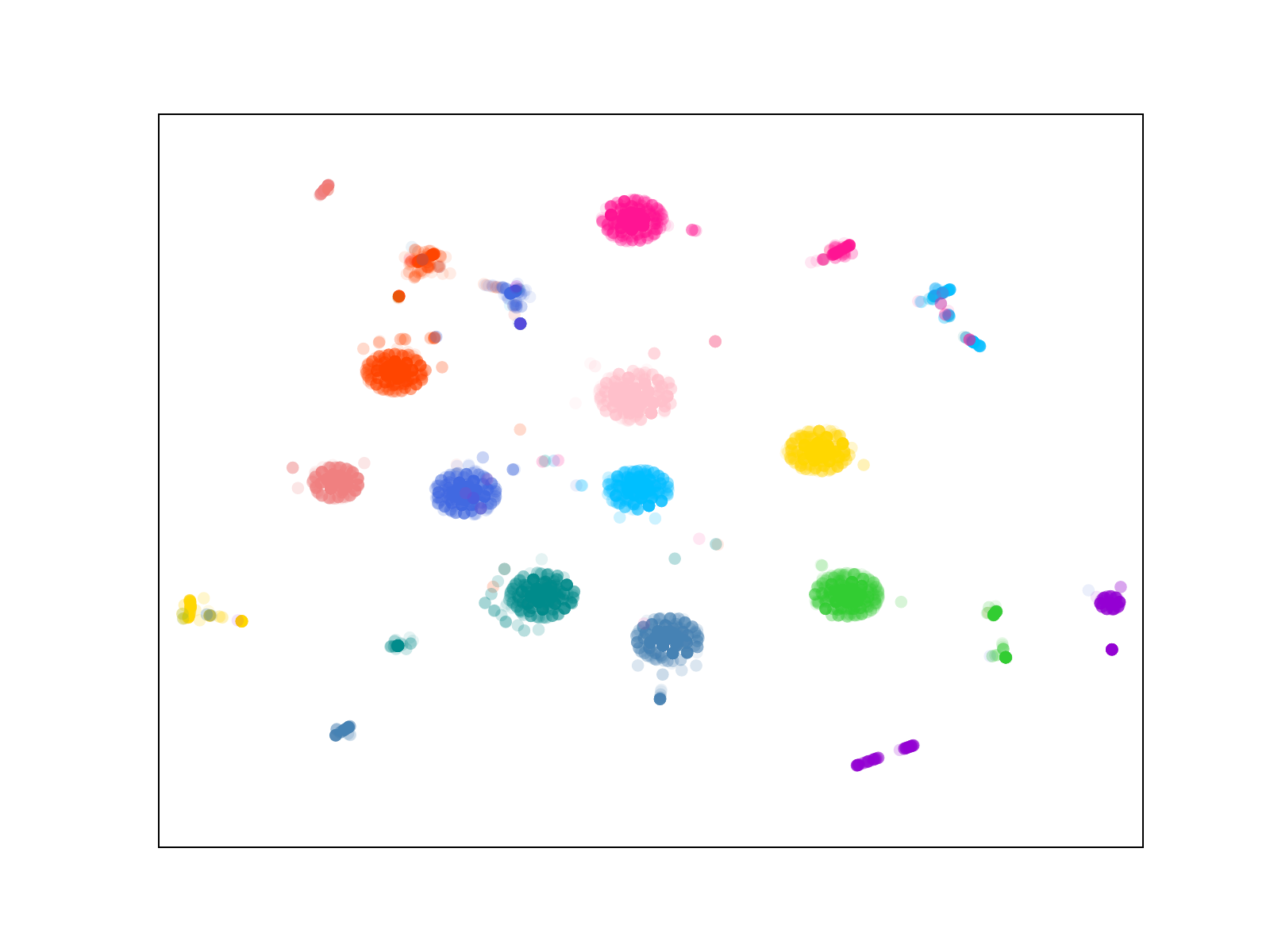}}
\caption{$\mathrm{BN}\oplus\mathrm{i}\textnormal{-}\mathrm{vect}\oplus\mathrm{GMMD}$}
\label{gmmd_bn_ivect_cons2}
\end{subfigure}
\end{center}
\caption{Analysis of lattice-based features for \textbf{\textit{consonants-2}} using t-SNE for TDNN models trained on different basic features.}
\label{fig_tsne_c2}
\end{figure}

First we analyzed the adaptation algorithm using t-SNE (Section~\ref{sec_tsne}). 
The results of the visual t-SNE analysis are given 
in Figure~\ref{fig_tsnev} for the group of vowels and 
in Figures \ref{fig_tsne_c1}, \ref{fig_tsne_c2} -- for the two group of consonants.
We can observe for all groups of phonemes that the adapted features 
(Figures~\ref{bn_ivect_vowels}, \ref{bn_ivect_cons1}, \ref{bn_ivect_cons2})
form  more distinct and clear phone clusters
than the unadapted features 
(Figures~\ref{mfcc_vowels}, \ref{mfcc_cons1}, \ref{mfcc_cons2}).
Also we can note the use of GMMD features helps to further slightly improve cluster separability
(Figures~\ref{gmmd_bn_ivect_vowels}, \ref{gmmd_bn_ivect_cons1}, \ref{gmmd_bn_ivect_cons2}).

\begin{table}
\caption{\it{Davies-Bouldin (DB) index for different types of features used in TDNN training. The DB index is calculated on lattice-based features produced by the corresponding model.}}\label{tab_db_index}
\centerline{
		\begin{tabular}{|c|c|c|c|}
		\hline
		Features  &   State 0 & State 1 & State 2 \\ \hline
         MFCC   & 1.67  & 1.52	& 1.71 \\       $\mathrm{BN}\oplus\mathrm{i}\textnormal{-}\mathrm{vectors}$    & 1.53  & 1.36	& 1.41 \\ 
$\mathrm{BN}\oplus\mathrm{i}\textnormal{-}\mathrm{vectors}\oplus\mathrm{GMMD}$   & \textbf{1.39}  & \textbf{1.26}	& \textbf{1.27 }\\ \hline
\end{tabular}}
\end{table}

\unboldmath{}
\begin{table}
\caption{\it{Statistics for lattice-based features, produced by the corresponding TDNN models. All  statistics in the table are calculated only for speech frames (excluding silence). The average log-probability of the correct phoneme is given with the standard deviation.}}\label{tab_stat}
\centerline{
\begin{tabular}{|c|c|c|c|}
\hline
Features  &    FER  & Oracle  FER & Aver. correct log-prob. \\ \hline
 MFCC   & 5.18  & \textbf{0.72}	& $-0.17\pm0.83$ \\       $\mathrm{BN}\oplus\mathrm{i}\textnormal{-}\mathrm{vectors}$    & 4.11  & 0.75	& $-0.11 \pm0.64$ \\ 
$\mathrm{BN}\oplus\mathrm{i}\textnormal{-}\mathrm{vectors}\oplus\mathrm{GMMD}$   & \textbf{3.64}  & 1.23 & $-0.08\pm0.52$ \\ \hline
\end{tabular}}
\end{table}

\unboldmath{}

To support this  visual analysis of cluster separation, we calculated 
DB index (Section~\ref{sec_db_inex}) for all phonemes, separately for each state type, depending on its position in phoneme HMM (State 0, 1, 2).
As we can see in Table~\ref{tab_db_index}, DB index decreases for all HMM states when we move from
unadapted (MFCC) to adapted ($\mathrm{BN}\oplus\mathrm{i}\textnormal{-}\mathrm{vectors}$) features.
That confirms the fact that the clusters are better for adapted features. The acoustic model with the adapted GMMD features ($\mathrm{BN}\oplus\mathrm{i}\textnormal{-}\mathrm{vectors}\oplus\mathrm{GMMD}$) shows the best result (the smallest value of DB index).

\begin{figure}
\begin{center}
\begin{subfigure}{0.47\textwidth}
\center{\includegraphics[width=57mm, trim = 0.2in 0.2in 0.2in 0.1in]{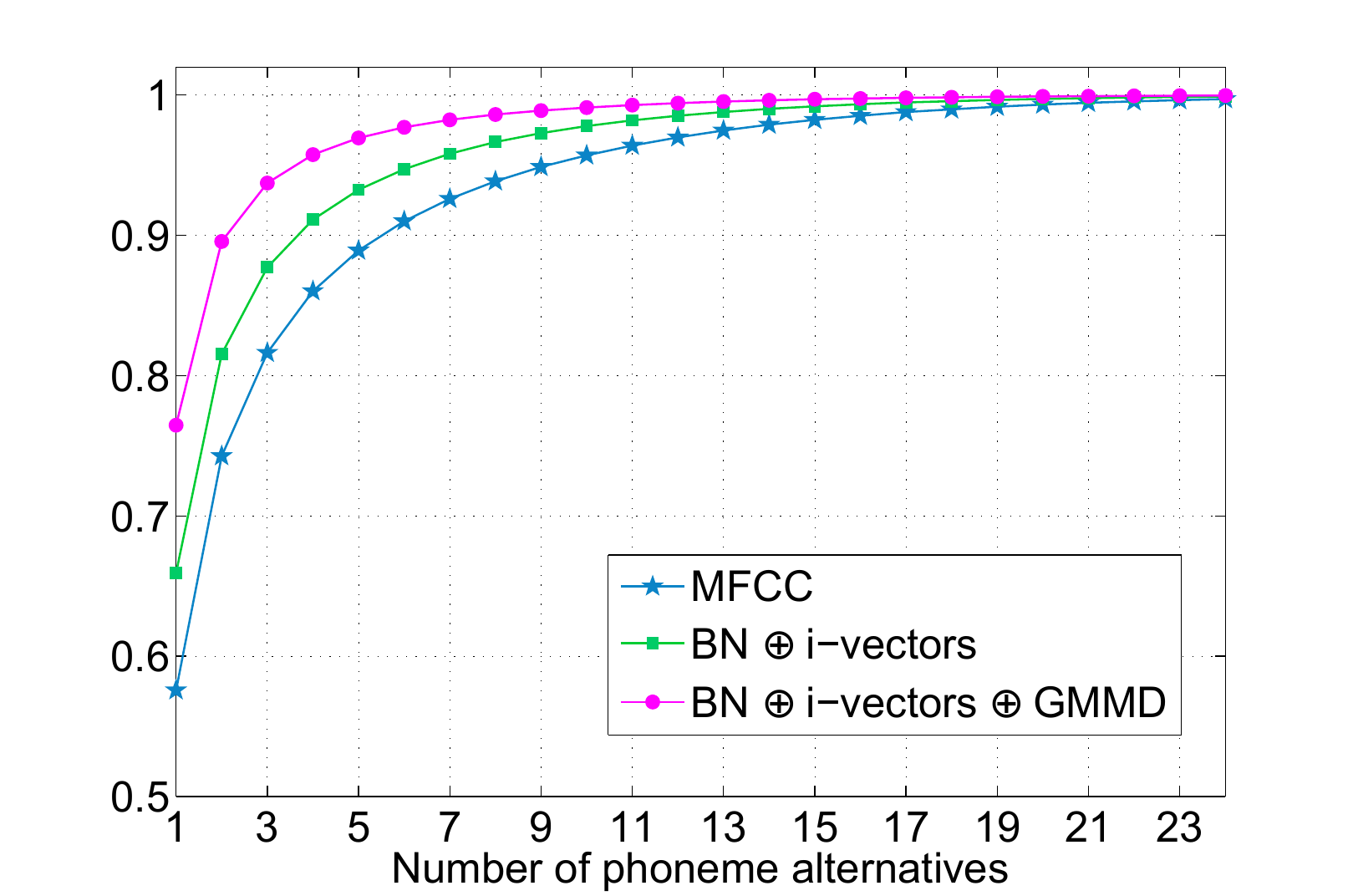}}
\caption{}
\label{size-row}
\end{subfigure}
\begin{subfigure}{0.47\textwidth}
\center{\includegraphics[width=57mm, trim = 0.2in 0.2in 0.2in 0.1in]{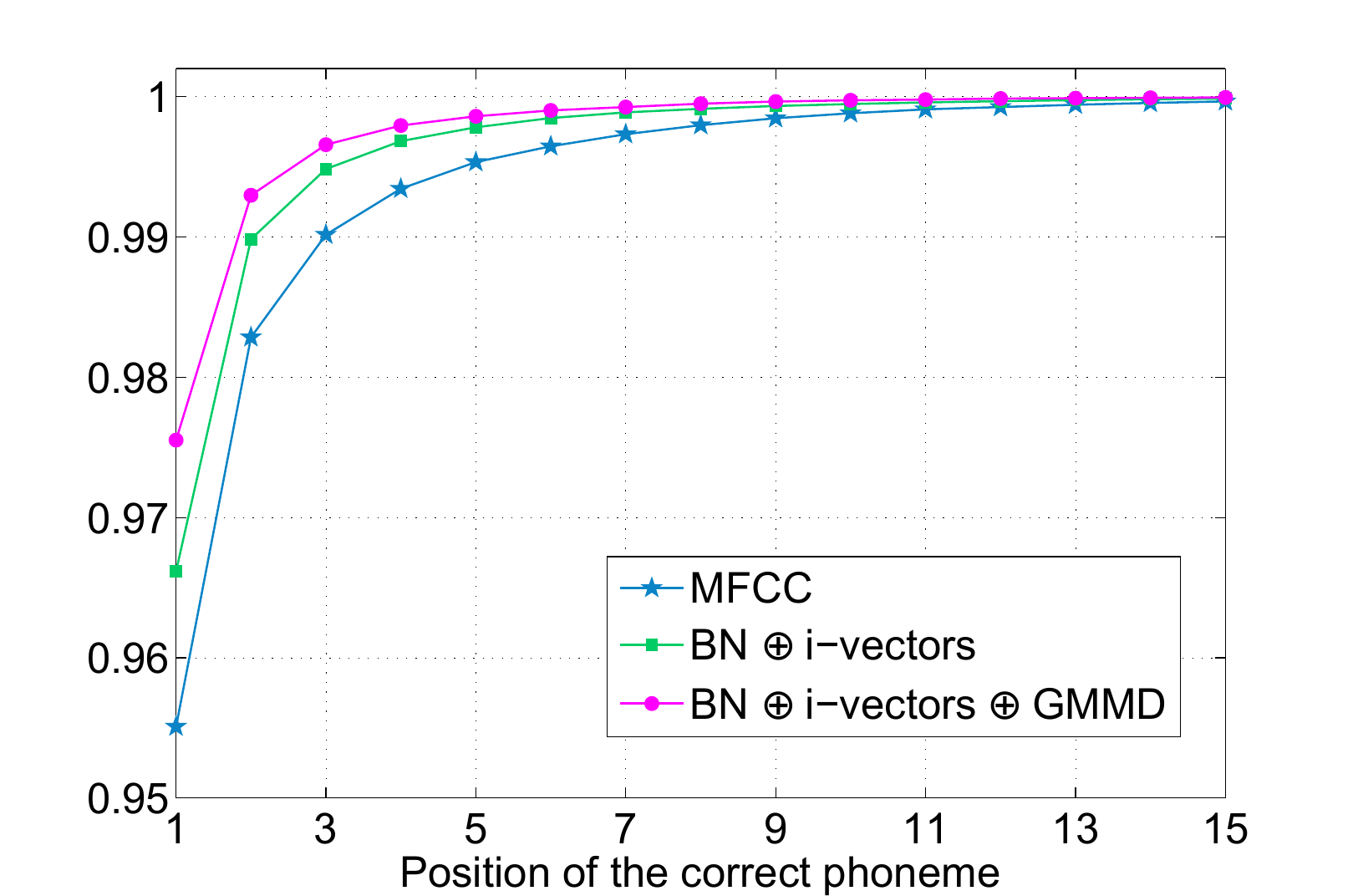}}
\caption{}
\label{position}
\end{subfigure}
\begin{subfigure}{0.47\textwidth}
\includegraphics[width=57mm, trim = 0.2in 0.2in 0.2in 0.1in]{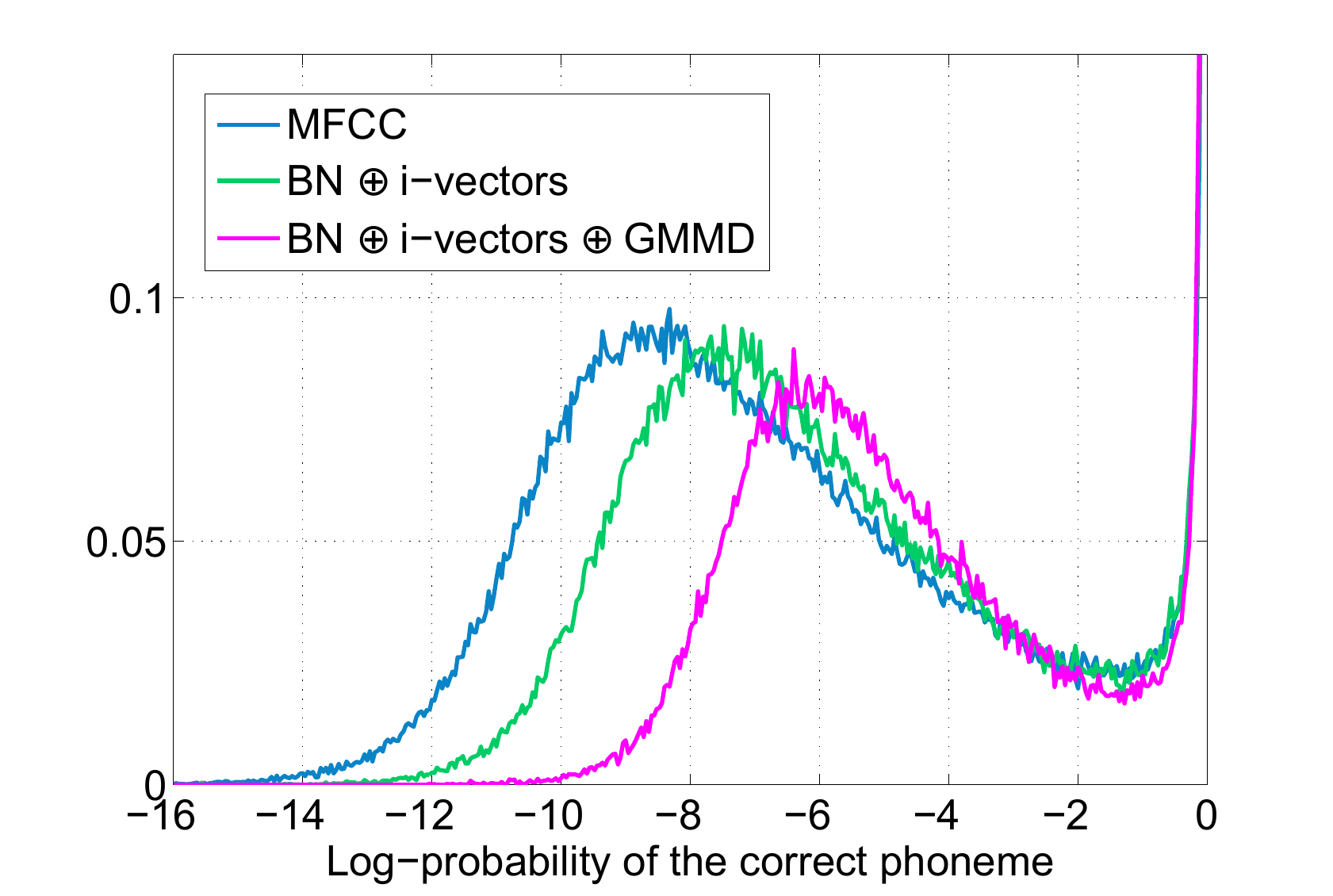}
\caption{}
\label{probs}
\end{subfigure}
\end{center}
\caption{Adaptation analysis based on lattice-based features statistics for the free TDNN models: 
(a) CDF  of the number of phoneme alternatives in the lattices for a certain frame;
(b) CDF  of the position of the correct phoneme in the list of all phoneme candidates (ordered by the posterior probability) presented in lattices for a certain frame;
(c) Log-probability histogram of the correct phoneme (if it exists) in the lattice for a certain frame.
}
\label{fig_hists}
\end{figure}

In order to more deeply investigate the adaptation behavior, we calculated additional statistics for lattices-based features (Table~\ref{tab_stat}). 
Frame error rate (FER) is calculated on the phoneme level using only  speech frames (excluding silence). 
Oracle FER was calculated also only on  speech frames as follows: if the correct phoneme was not present in the list of all candidates in the lattices for a given frame, then it was considered as an error. 

We can see that FER decreases when moving from the unadapted features to the adapted ones, and then to the use of the adapted GMMD features, that correlates with the WER behavior (Table~\ref{table_results_tdnn_adapt}). It is interesting to note, that Oracle FER, on the contrary, increases with the adaptation. 
One of the possible explanation for this unusual situation can be phonetic transcription errors which occur in the lexicon. 
The  adapted models, which can be more sensitive to the phonetic transcription errors, can more strictly supplant, during the decoding process,  hypotheses  
that do not match the acoustic signal.\footnote{For GMM-HMM models, it is known~\cite{gollan2008confidence}  that MAP adaptation can reinforce  errors present in  transcriptions used for adaptation.}

Decoding parameters, such as \textit{decoding beam} and \textit{lattice beam}, were the same for all models, but the adapted models in average have less alternative phoneme candidates for a certain frame, than the unadapted one.
This can bee seen in  Figure~\ref{size-row}, which shows the \textit{cumulative distribution functions} (CDFs)  of the number of phoneme alternatives presented in the lattices for a certain frame, estimated only for speech frames. 
Figure~\ref{position} demonstrates CDFs of position of the correct phoneme (if it exists) in lattices for a certain speech frame in the list of all phoneme candidates ordered by their posterior probabilities.
We can conclude from this figure that for adapted models (especially for the AM with GMMD features), the correct candidate has less incorrect alternatives with higher probabilities that its own.

Also, the average \textit{correct log-probability} (it is a value from a lattice based features vector, which corresponds to the correct phoneme for a given frame) has a maximum value for \boldmath{$\mathrm{TDNN}_{\mathrm{BN}\oplus\mathrm{i}\textnormal{-}\mathrm{vectors}\oplus\mathrm{GMMD}}$} model (see the last column of Table~\ref{table_results_tdnn_adapt} 
and a histogram on Figure~\ref{probs}).

Hence, if we compare the statistics presented in Table~\ref{tab_stat} and in Figure~\ref{fig_hists}, we can conclude that 
the adapted models tend to be more "selective" and "discriminative" in comparison with
unadapted models in the sense that:
(1) they reduce the number of alternatives in the hypothesis more aggressively;
(2) they give higher probability values for correct candidates;
and (3) the correct phoneme candidate, if it exists in the lattice for a given frame, has in average, less incorrect competitor alternatives with higher probabilities than its own.
The AM trained on the adapted GMMD features most  strongly  shows
the same properties.\footnote{In \cite{woodland2001speaker,woodland1996iterative} for GMM-HMM acoustic models and MLLR adaptation, it was also noticed  that adaptation allows to obtain smaller and more accurate lattices.}

This analysis demonstrates that the acoustic models trained with the proposed adapted GMMD features perform better than the baseline adapted model not only by comparing  the WER  (which is the main metric), but also on the other levels.

Also, what is important, this gives us an understanding of the possible way 
of the adaptation performance improvement 
through more careful handling of the transcripts, for example,
by automatically estimation their quality and reliability.

\unboldmath{}
\begin{figure}[t]
\begin{center}
\begin{subfigure}{0.48\textwidth}
\center{\includegraphics[width=63mm, trim = 0.4in 0.2in 0.1in 0.5in]{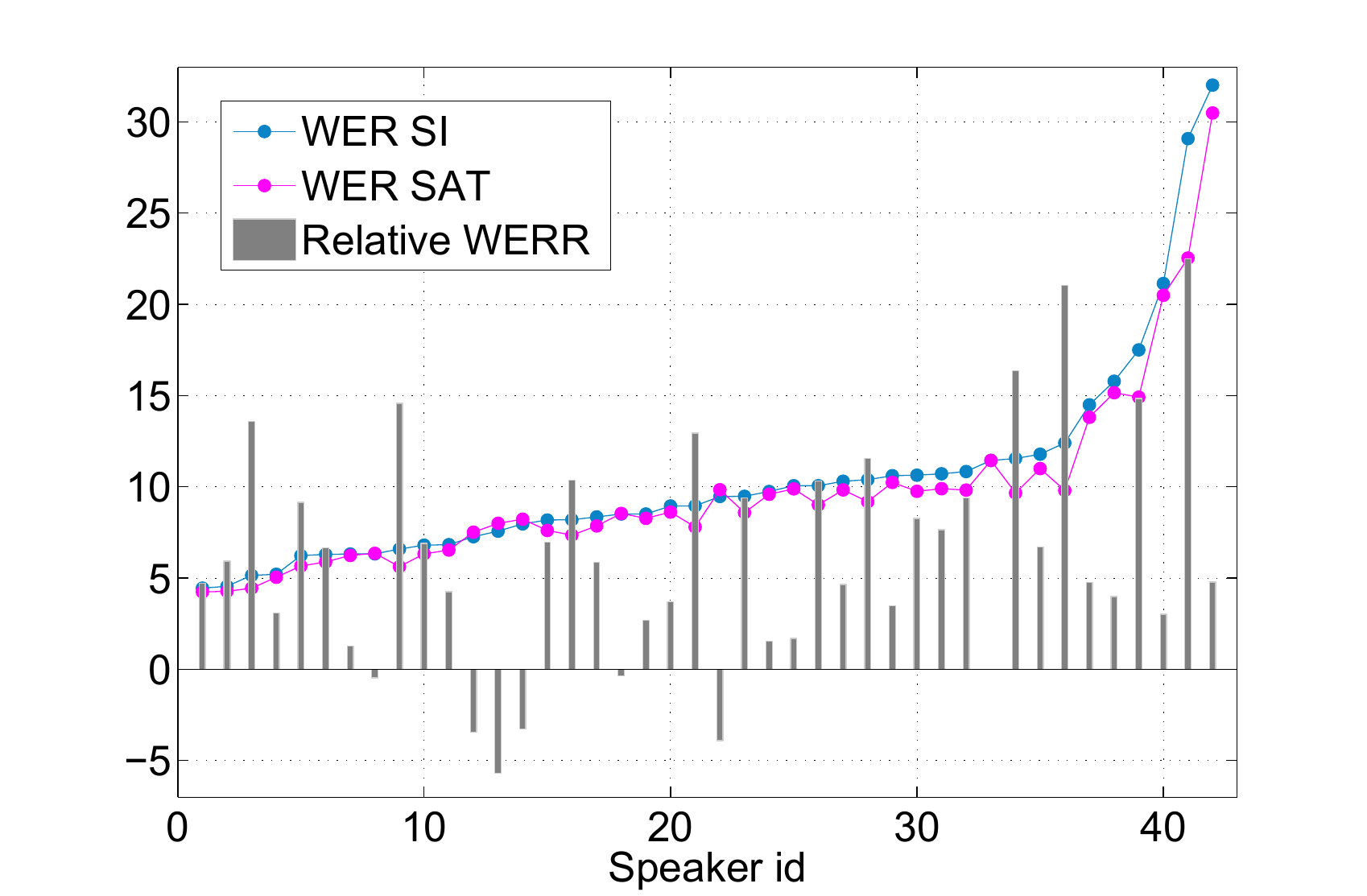}}
\caption{}
\label{DeltaWER}
\end{subfigure}
\begin{subfigure}{0.48\textwidth}
\center{\includegraphics[width=61mm, trim = 0.1in 0.1in 0.1in 0.1in]{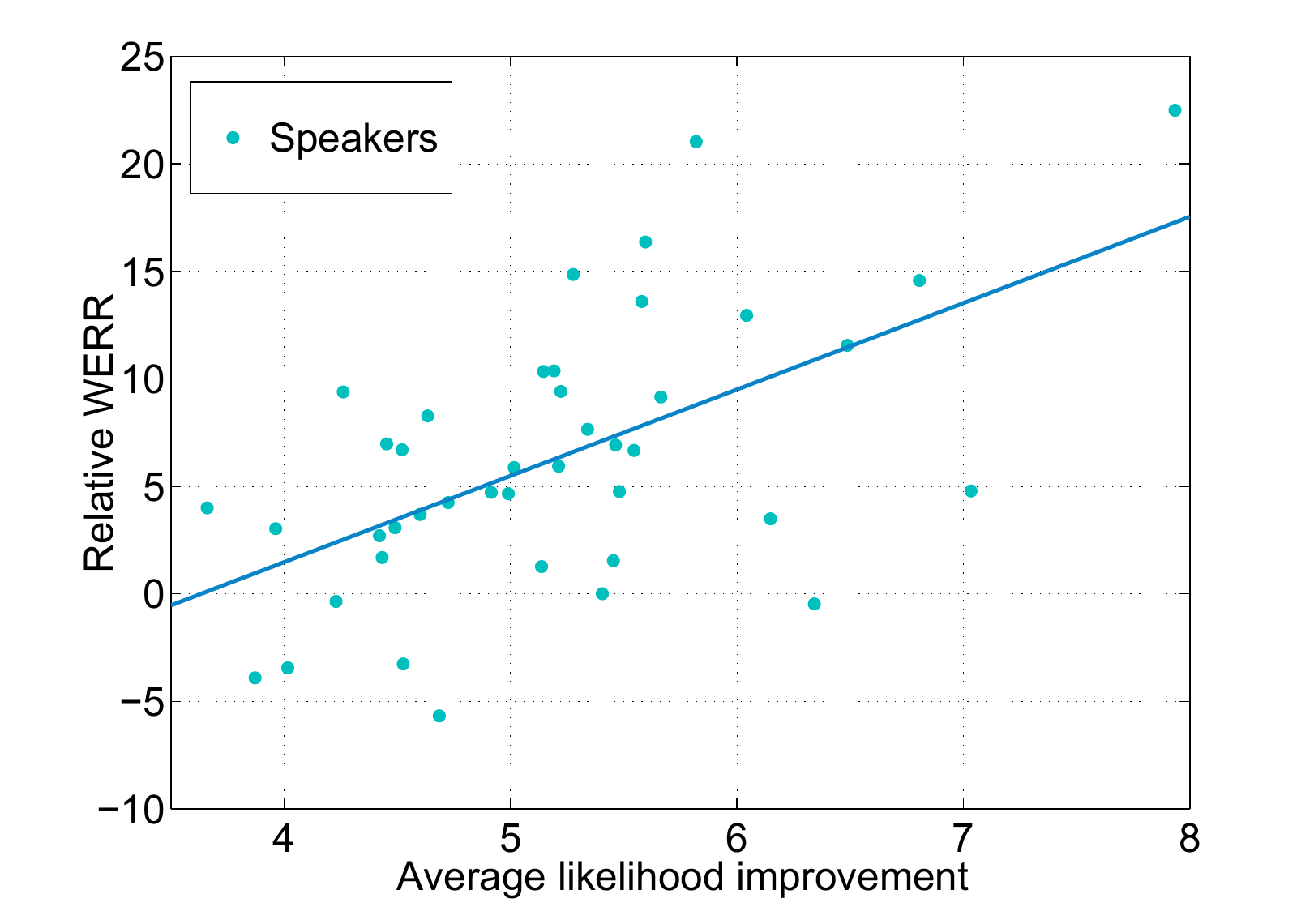}}
\caption{}
\label{LikeImprWERR}
\end{subfigure}
\caption{Summary statistics for all speakers from the development and the two test data sets: 
(a) WERs(\%) for two (SI and SAT) TDNN models:
SI -- \boldmath{$\mathrm{TDNN}_{\mathrm{BN}}$}, 
SAT -- \boldmath{$\mathrm{TDNN}_{\mathrm{BN}\oplus\mathrm{GMMD}}$}.
Relative WER reduction (Relative WERR,\%) is computed for the given WERs. Results are ordered by increasing WER values for the SI model; 
(b)  Dependence of relative WERR (the same as in (a)) on average likelihood improvement, obtained by MAP adaptation of the auxiliary monophone model. The line corresponds to the linear regression model.}
\label{fig_byspeakers}
\end{center}
\end{figure}

\unboldmath{}
In addition, Figure~\ref{fig_byspeakers} shows the statistics obtained for 42 speakers from $Develop\-ment$, $Test_{1}$ and $Test_{2}$ data sets for 
\boldmath{$\mathrm{TDNN}_{\mathrm{BN}}$}, 
\boldmath{$\mathrm{TDNN}_{\mathrm{BN}\oplus\mathrm{GMMD}}$}
models.
\unboldmath{}
We can observe in Figure~\ref{DeltaWER} that the proposed adaptation approach improves recognition accuracy for 83\% of speakers.
For the same speakers, Figure~\ref{LikeImprWERR} illustrates the dependence of relative WER reduction from average likelihood improvement, obtained by MAP adaptation of the auxiliary monophone model.

\begin{figure}
\begin{center}
\includegraphics[width=80mm, trim = 0.1in 0.1in 0.1in 0.1in]{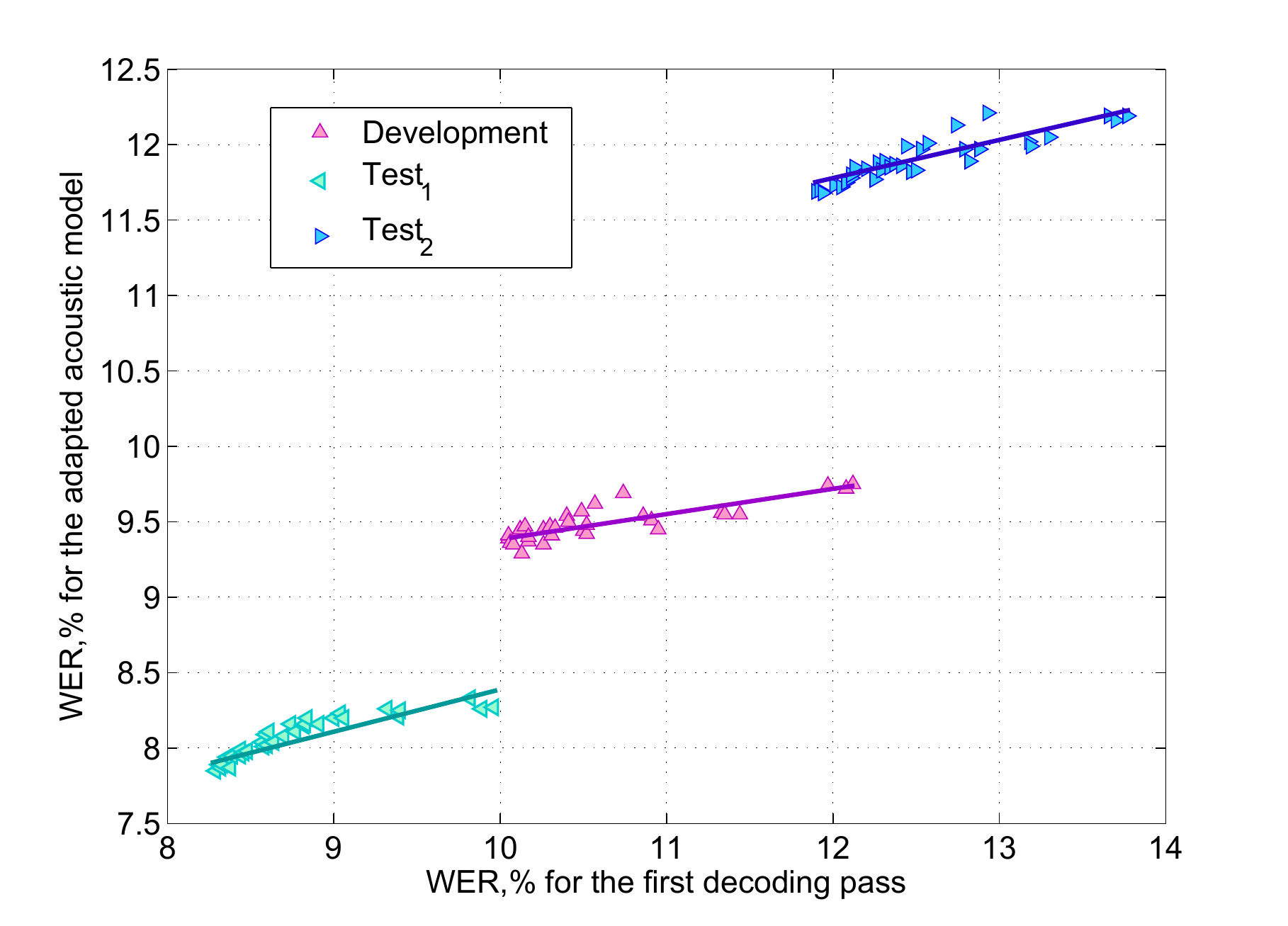}
\caption{Dependence of WER,\% (for the adapted model \boldmath{$\mathrm{TDNN}_{\mathrm{BN}\oplus\mathrm{i}\textnormal{-}\mathrm{vectors}\oplus\mathrm{GMMD}}$}) from the quality  of the adaptation targets for the development and the two test data sets.}
\label{fig_depend_on_wer}
\end{center}
\end{figure}

Finally, we explored how sensitive the  proposed adaptation is to the 
quality of the transcriptions from the first decoding pass.
This is an important aspect of  adaptation algorithms.
For GMM-HMM models, it is known, that transform-based adaptation approaches, limited in the number of
free adaptation parameters, are robust transcription errors~\cite{gollan2008confidence}. For DNN unsupervised adaptation, this problem was investigated for LHUC adaptation in~\cite{Swietojanski2016learning} where it was concluded that the LHUC algorithm is not very sensitive to the quality of adaptation targets.
In our study, we varied (degraded) the quality of the transcriptions used in adaptation by using sub-optimal decoding parameters (\textit{word insertion penalty} and
\textit{language model weight})
in the first decoding pass and performed  adaptation of \boldmath{$\mathrm{TDNN}_{\mathrm{BN}\oplus\mathrm{i}\textnormal{-}\mathrm{vectors}\oplus\mathrm{GMMD}}$}
using these different targets. 
The results of this experiment are presented in Figure~\ref{fig_depend_on_wer}.
We can observe  that  changes in WERs in the first decoding pass lead to the changes in
the quality of the adapted AM. However,  these changes are not so dramatic.

\unboldmath{}
\section{Conclusions}\label{Sect_Concl}
In this paper we have investigated the GMM framework for adaptation of DNN-HMM acoustic models
and combination of MAP-adapted GMM-derived with conventional features at different levels of DNN and TDNN architectures.

Experimental results on the TED-LIUM corpus demonstrate that, in an unsupervised adaptation mode, the proposed adaptation and fusion techniques can provide approximately,
a $12$--$18\%$  relative WERR on different adaptation sets, compared to the SI DNN system built on conventional features,  
and a $4$--$6\%$ relative WERR compared to the  strong adapted baseline --- SAT-DNN trained on fMLLR adapted features.
For TDNN models using the adapted GMMD features and fusion techniques leads to improvement of $10$--$26\%$ WERR in comparison with SI model trained on conventional features and 
$7$--$13\%$ WERR in comparison with SAT model trained with i-vectors.
Hence, for  both considered adaptation techniques, fMLLR and i-vectors, the proposed adaptation approach has appeared to be complementary and provide an additional improvement in recognition accuracy.

We have looked from the various points of view at the proposed adaptation approach exploring the latticed-based features, generated from the decoding lattices and have demonstrated, that the advantage of using MAP-adapted  GMMD features manifests itself at different levels of the decoding process. 
This analysis also shows a possible potential and direction for improvement of the proposed adaptation approach through more careful handling of quality of the phonetics transcripts, used in adaptation.
This will be a focus of our future work.

\bibliography{mybibfile}

\appendix

\section{Implementation details for baseline DNN models}\label{app1}
This appendix illustrates the details of our experiments with DNN models.

We trained four baseline DNN acoustic models:
\begin{itemize}
\item \boldmath{$\mathrm{DNN}_\mathrm{BN}\textnormal{-}\mathrm{CE}$} 
was trained on BN features with CE criterion.
\item \boldmath{$\mathrm{DNN}_\mathrm{BN}\textnormal{-}\mathrm{sMBR}$} 
was obtained from the previous one by performing four epochs of sequence-discriminative training with  Minimum Bayes Risk (sMBR) criterion.
\item $\mathrm{DNN}_{\mathrm{BN\textnormal{-}fMLLR}}\textnormal{-}\mathrm{CE}$
was trained on fMLLR-adapted BN features.
\item $\mathrm{DNN}_{\mathrm{BN\textnormal{-}fMLLR}}\textnormal{-}\mathrm{sMBR}$
was obtained from the previous one by four epochs of sMBR sequence-discriminative training.
\end{itemize}

For training DNN models, the initial GMM model was trained  using
39-dimensional MFCC features including delta and acceleration coefficients. 
Linear discriminant analysis (LDA), followed by maximum likelihood linear transform (MLLT) and then fMLLR transformation, was then applied over these MFCC features to build a GMM-HMM system. 
Discriminative training with the boosted maximum mutual information (BMMI) objective  was finally performed on top of this model.

Then a DNN was trained for BN feature extraction. The DNN system was trained using the frame-level cross entropy criterion and the senone (tied-state) alignment generated by the GMM-HMM system.
To train this DNN, 40-dimensional log-scale filterbank features concatenated with 3-dimensional pitch-features were spliced across 11 neighboring frames, resulting in 473-dimensional $(43\times11)$ feature vectors. After that a DCT transform was applied and the dimension was reduced to 258. A DNN model for extraction 40-dimensional BN features was trained with the following topology: one 258-dimensional input layer; four hidden layers (HL), where the third HL was a BN layer with 40 neurons and other three HLs were 1500-dimensional; the output layer was 2390-dimensional.
Based on the obtained BN features we trained the GMM model, which was used to produce the forced alignment, and then SAT-GMM model was trained on
 fMLLR-adapted BN features. 
 Then fMLLR-adapted BN features were spliced in time with the context of 13 frames: [-10,-5..5,10] to train the final DNN model.
The final DNN had a 520-dimensional input layer; six 2048-dimensional HLs with logistic sigmoid activation function, and a 4184-dimensional softmax output  layer, with units corresponding to the context-dependent states. 

The DNN parameters were initialized with stacked restricted Boltzmann machines (RBMs) by using layer by layer generative pre-training. It was trained with an initial learning rate of 0.008 using the
cross-entropy objective function to obtain the SAT-DNN-CE model
\boldmath{}$\mathrm{DNN}_{\mathrm{BN\textnormal{-}fMLLR}}\textnormal{-}\mathrm{CE}$.

After that, four epochs of sequence-discriminative training with per-utterance updates, optimizing state
sMBR criteria, were performed to obtain the SAT-DNN-sMBR model 
$\mathrm{DNN}_{\mathrm{BN\textnormal{-}fMLLR}}\textnormal{-}\mathrm{sMBR}$.

Baseline SI DNN models 
(\boldmath{$\mathrm{DNN}_\mathrm{BN}\textnormal{-}\mathrm{CE}$} 
and
\boldmath{$\mathrm{DNN}_\mathrm{BN}\textnormal{-}\mathrm{sMBR}$})
were trained in a similar way as the SAT DNNs described above, but without fMLLR adaptation. 
\unboldmath{}

\section{Implementation details for baseline TDNN models}\label{app2}
This appendix illustrates the details of our experiments with TDNN models.

We trained four baseline TDNN acoustic models, which differ only in the type of the input features:
\begin{itemize}
\item \boldmath{$\mathrm{TDNN}_\mathrm{MFCC}$} 
was trained on high-resolution MFCC features.
\item \boldmath{$\mathrm{TDNN}_{\mathrm{MFCC}\oplus\mathrm{i}\textnormal{-}\mathrm{vectors}}$}
was trained on high-resolution MFCC features, appended with 100-dimensional i-vectors.
\item \boldmath{$\mathrm{TDNN}_{\mathrm{BN}}$}
was trained on BN features.
\item \boldmath{$\mathrm{TDNN}_{\mathrm{BN}\oplus\mathrm{i}\textnormal{-}\mathrm{vectors}}$}
was trained on BN features, appended with 100-dimensional i-vectors.
\end{itemize}
The baseline SAT-TDNN model \boldmath{$\mathrm{TDNN}_{\mathrm{MFCC}\oplus\mathrm{i}\textnormal{-}\mathrm{vectors}}$}    is similar to those described in~\cite{peddinti2015time}, except for the number of hidden layers and slightly different subsequences of splicing and sub-sampling indexes. The two types of data augmentation strategies were  applied for the speech  training data:  speed perturbation (with factors 0.9, 1.0, 1.1) and volume perturbation. 
The SAT-TDNN model was trained on high-resolution MFCC features (without dimensionality reduction, keeping all
40 cepstra) concatenated with  100-dimensional i-vectors. The temporal context was 
\unboldmath{$[t-16, t+12]${} and the splicing indexes
used here were 
\unboldmath{$[-2,2]$, $\{-1,2\}$, $\{-3,3\}$, $\{-7,2\}$, $\{0\}$, $\{0\}$}}. 
This model has 
850-dimensional hidden layers with rectified linear units (ReLU)~\cite{Dahl2013improving} activation functions, a 4052-dimensional output layer and 
approximately 10.9 million  parameters.

The baseline SI-TDNN model  \boldmath{$\mathrm{TDNN}_\mathrm{MFCC}$}  was trained in a similar way as the SAT-TDNN described above, but without using i-vectors. 

In addition to these baseline models, for comparison purpose, we trained two other baseline TDNNs (with and without i-vectors) using BN features instead of high-resolution MFCC features. The same BN features we used later for training an auxiliary monophone GMM model for GMMD feature extraction. These BN features were extracted using a DNN trained in a similar way as described in Section~\ref{sec_base} for DNN AM, but on the  high-resolution MFCC features (instead of "filter bank $\oplus$ pith" features) and on the augmented (by means of speed and volume perturbation) data base. As in~\cite{peddinti2015time}, the i-vectors during the training were calculated  every two utterances.

\section{Implementation details for DNNs trained on GMMD features}\label{app3}
This appendix presents the details of training DNN models on the proposed GMMD features.
In this set of experiments we trained four DNNs, using the approach proposed in Section~\ref{DNN-GMMD}:

\begin{itemize}
\item \boldmath{$\mathrm{DNN}_{\mathrm{GMMD}\oplus\mathrm{BN}}\textnormal{-}\mathrm{CE}$}
is a DNN without performing speaker adaptive training, which in our case means that the auxiliary GMM monophone model was not adapted. This DNN model was trained using the CE criterion.
\item \boldmath{$\mathrm{DNN}_{\mathrm{GMMD}\oplus\mathrm{BN}}\textnormal{-}\mathrm{sMBR}$}
was obtained from the previous one by performing four epochs of sequence-discriminative training with per-utterance updates, optimizing the sMBR criterion.
\item \boldmath{
$\mathrm{DNN}_{\mathrm{\mathrm{GMMD\textnormal{-}MAP}\oplus\mathrm{BN}}}\textnormal{-}\mathrm{CE}$}
was a proposed SAT DNN model trained on speaker adapted GMMD-MAP features, with the CE criterion.
\item 
$\mathrm{DNN}_{\mathrm{\mathrm{GMMD\textnormal{-}MAP}\oplus\mathrm{BN}}}\textnormal{-}\mathrm{sMBR}$
was obtained from the previous one by performing four epochs of sMBR sequence training.
\end{itemize}

Models
$\mathrm{DNN}_{\mathrm{\mathrm{GMMD\textnormal{-}MAP}\oplus\mathrm{BN}}}\textnormal{-}\mathrm{CE}$
and 
$\mathrm{DNN}_{\mathrm{GMMD\textnormal{-}MAP}\oplus\mathrm{BN}}\textnormal{-}\mathrm{sMBR}$
were trained as described in Section~\ref{DNN-GMMD}.
The GMMD features were extracted using a monophone auxiliary GMM model, trained on BN features. This GMM model was adapted for each speaker by MAP adaptation algorithm (Section~\ref{sec_map}).

We took the state tying from the baseline SAT-DNN to train all other models.
The purpose of using the same state tying is to allow posterior level fusion for these models. 
We also took an alignment obtained by the SAT-DNN model, because in~\cite{Tomashenko2016anew} it was shown to slightly improve the result.
 
Then we concatenated two types of BN features and spliced them for training the final DNN model.
Both DNN models were trained on the proposed features in the same manner and had the same topology except for the input features, as the final baseline SAT DNN model trained on BN features
(Section \ref{sec_base}). 
The other two SI models (\boldmath{$\mathrm{DNN}_{\mathrm{GMMD}\oplus\mathrm{BN}}\textnormal{-}\mathrm{CE}$} and \boldmath{$\mathrm{DNN}_{\mathrm{GMMD}\oplus\mathrm{BN}}\textnormal{-}\mathrm{sMBR}$}) were trained in the same manner but without speaker adaptation.

\end{document}